\documentclass[aps,prl,twocolumn,superscriptaddress,showpacs]{revtex4-2}
\usepackage{amsmath}
\usepackage{color}
\usepackage{amssymb}
\usepackage{epsfig}
\usepackage{multirow}
\usepackage{amsbsy}
\usepackage{array}
\usepackage{diagbox}
\usepackage{bm}
\usepackage{extarrows}
\usepackage{graphicx}
\usepackage{appendix}
\usepackage[normalem]{ulem}
\usepackage{ulem}
\usepackage{txfonts}
\usepackage[T1]{fontenc}
\usepackage{booktabs}
\allowdisplaybreaks[1]

\newcommand{\bk}{\mathbf{k}}
\newcommand{\br}{\mathbf{r}}

\newcommand{\bq}{\mathbf{q}}
\newcommand{\ham}{\mathcal{H}}

\usepackage[colorlinks=true,linkcolor=blue,citecolor=blue,urlcolor=blue,bookmarks=false]{hyperref}

\begin{document}
	
\title{Interplay between Pair Density Wave and a Nested Fermi Surface}

\author{Jin-Tao Jin}
\affiliation{Kavli Institute for Theoretical Sciences, University of Chinese Academy of Sciences, Beijing 100190, China}

\author{Kun Jiang}
\email{jiangkun@iphy.ac.cn}
\affiliation{Institute of Physics, Chinese Academy of Sciences, Beijing 100190, China}
\affiliation{Songshan Lake Materials Laboratory, Dongguan, Guangdong 523808, China}

\author{Hong Yao}
\email{yaohong@tsinghua.edu.cn}
\affiliation{Institute for Advanced Study, Tsinghua University, Beijing 100084, China}
\affiliation{State Key Laboratory of Low Dimensional Quantum Physics, Tsinghua University, Beijing 100084, China}

\author{Yi Zhou}
\email{yizhou@iphy.ac.cn}
\affiliation {Institute of Physics, Chinese Academy of Sciences, Beijing 100190, China}
\affiliation{Songshan Lake Materials Laboratory, Dongguan, Guangdong 523808, China}
\affiliation{Kavli Institute for Theoretical Sciences, University of Chinese Academy of Sciences, Beijing 100190, China}
\affiliation{CAS Center for Excellence in Topological Quantum Computation, University of Chinese Academy of Sciences, Beijing 100190, China}

\date{\today}
\begin{abstract}
We show that spontaneous time-reversal-symmetry (TRS) breaking can naturally arise from the interplay between pair density wave (PDW) ordering at multiple momenta and nesting of Fermi surfaces (FS).
Concretely, we consider the PDW superconductivity on a hexagonal lattice with nested FS at $3/4$ electron filling, which is related to a recently discovered superconductor CsV$_3$Sb$_5$. Because of nesting of the FS, each momentum $\bk$ on the FS has at least two counterparts $-\bk\pm\mathbf{Q}_{\alpha}$ ($\alpha=1,2,3$) on the FS to form finite momentum ($\pm\mathbf{Q}_{\alpha}$) Cooper pairs, resulting in a TRS and inversion broken PDW state with stable Bogoliubov Fermi pockets.  
Various spectra, including (local) density of states, electron spectral function and the effect of quasi-particle interference, have been investigated. The partial melting of the PDW will give rise to $4\times{}4$ and $\frac{4}{\sqrt{3}}\times\frac{4}{\sqrt{3}}$ charge density wave (CDW) orders, in addition to the $2\times2$ CDW. 
Possible implications to real materials such as CsV$_3$Sb$_5$ and future experiments have been further discussed.
\end{abstract}

\maketitle

{\em Introduction. ---}
A pair density wave (PDW) is a superconducting (SC) state in which 
Cooper pairs carry finite momentum and its SC order parameter is spatially modulated without 
external magnetic field~\cite{PDW-Kivelson2007,Agterberg2008NPdislocations,Berg-cuprate-NJP2009,Berg-PRL2010,Fradkin-PRB2012,cho2012superconductivity,M-F_model_2,PALee-PRX2014,maciejko2014weyl,YXWang-PRL2015,SKJian-PRL2015,SKJian-PRL2017,YXWang-PRB2018,Kim-SciAdv2019,ZYHan-PRL2020,LFu-PRB2020,zhou2021chern,Huang-npjQM2022,Yao-arX2022} 
(see, e.g. a recent review~\cite{PDW}).
Such kind of state is similar to the one proposed earlier by Fulde-Ferrell (FF)~\cite{FF_state} and Larkin-Ovchinnikov (LO)~\cite{LO_state} (together known as FFLO) in magnetic field above the Pauli limit. 
As a mother state for various descendant orders, e.g., charge density wave (CDW), loop current, and charge-$4e$ superconductivity, PDW has been receiving increasing attentions from diverse fields in physics~\cite{PDW,RMP04}.
In particular, PDW was recently proposed as a promising candidate for explaining various interesting phenomena in cuprates and other strongly correlated systems~\cite{PDW,Hamidian2016,WangNP18,JCDavis19,WangPRX21}.

In previous studies of PDW,
only electrons near hot spots on Fermi surface (FS) can come into being finite momentum Cooper pairs and are gapped, while other parts of FS remain gapless.
In contrast, as will be revealed in this work, the FS nesting feature admits full PDW pairing around the FS, that will gain more condensation energy than the partial pairing usually, while in-gap quasi-particle excitations are still allowed. Thus, it would be of great interest to examine the interplay between these two, PDW and FS nesting. (Note that the role of FS nesting was considered mostly to CDW or spin-density-wave, e.g. see Ref.~\cite{Peierls,Frohlich54,CDW88,SDW60,SDW62,SDW96}). 

In this letter, we study PDW ordering in the presence of a nested FS 
on hexagonal lattices.
We found that a time reversal symmetry (TRS) breaking PDW state is energetically favored.
Bogoliubov quasi-particle excitations, density of states (DOS), 
local DOS (LDOS), and electron spectral function will be investigated as well as the quasi-particle interference (QPI) in scanning tunnelling microscopy (STM). 
The implications to recently discovered Kagome SC AV$_3$Sb$_5$(A=K,Rb,Cs) will be discussed.

\begin{figure}[tb]
	\includegraphics[width=\linewidth]{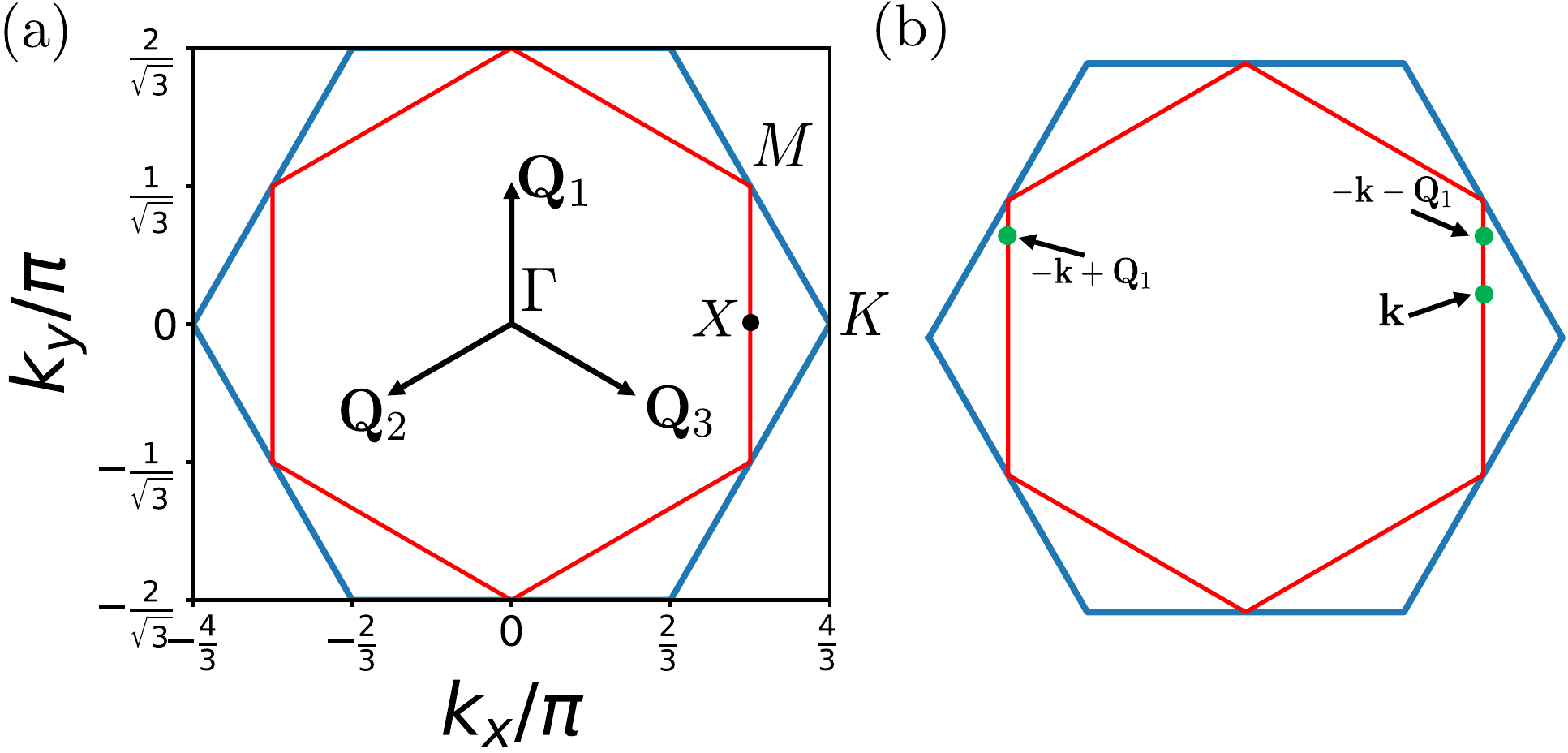}
	\caption{First Brillouin zone (BZ) of a hexagonal lattice and nesting feature of PDW. (a) BZ and the nested FS at $3/4$ filling. The boundary of BZ is in blue and red lines represent the FS. (b) Each $\bk$ on a FS segment along the $\mathbf{Q}_{\alpha}$ direction has two counterparts $-\bk\pm\mathbf{Q}_{\alpha}$ on the FS to form finite momentum ($\pm\mathbf{Q}_{\alpha}$) Cooper pairs.}\label{nesting}
\end{figure}

\emph{Model. ---}
We start with a single band model on a hexagonal lattice, on which the FS is nested as illustrated in Fig.~\ref{nesting}. The Hamiltonian takes the form:
\begin{equation}\label{H}
\begin{split}
H =  \sum_{\mathbf{k},\sigma}\xi_{\bk}c^{\dagger}_{\bk,\sigma}c_{\bk,\sigma}
+\sum_{\bk,\alpha} & \left[\Delta_{{\mathbf{Q}_\alpha}}(\bk)c^{\dagger}_{\bk,\uparrow}c^{\dagger}_{-\bk+\mathbf{Q}_\alpha,\downarrow}\right.\\
& + \left.\Delta_{-\mathbf{Q}_\alpha}(\bk)c^{\dagger}_{\bk,\uparrow}c^{\dagger}_{-\bk-\mathbf{Q}_\alpha,\downarrow}+
		h.c.
		\right],
\end{split}
\end{equation}	
where $c^{\dagger}_{\bk,\sigma}(c_{\bk,\sigma})$ is electron creation (annihilation) operator with momentum $\bk$ and spin $\sigma= \uparrow, \downarrow$, $\xi_\bk=\epsilon_\bk-\mu$ is the energy measured from the chemical potential $\mu$.  $\Delta_{\pm \mathbf{Q}_\alpha}(\bk) = \Delta_{\pm \mathbf{Q}_\alpha}\text{exp}\left[-(|\xi_{\bk}|+|\xi_{-\bk\pm\mathbf{Q}_\alpha}|)/(2\Lambda)\right]$ ($\alpha=1,2,3$) indicates the Cooper pairing with total momentum $\pm \mathbf{Q}_\alpha$. Here $\Lambda$ is an energy cutoff. Setting lattice constant $a=1$, we consider 
$\mathbf{Q}_1,\mathbf{Q}_2,\mathbf{Q}_3=\left(0,\frac{\pi}{\sqrt{3}}\right),\left(-\frac{\pi}{2},-\frac{\pi}{2\sqrt{3}}\right),\left(\frac{\pi}{2},-\frac{\pi}{2\sqrt{3}}\right)$, since they are most relevant to CsV$_3$Sb$_5$ for which experimental evidences of period 4 PDW was recently reported \cite{hongjun_gao}. 
As shown in Fig.~\ref{nesting}(a),
a hexagonal and nested FS 
can be realized by choosing 
$\mu$ or the electron filling properly.
As a simple example, we focus on a triangular lattice, set the nearest neighboring (NN) hopping integral $t=1$, and choose $\mu=2$ (or $3/4$ filling), such that {$\xi_\bk=-2\left[\cos\left(k_x\right)+\cos\left(\frac{1}{2}k_x+\frac{\sqrt{3}}{2}k_y\right)+\cos\left(\frac{1}{2}k_x-\frac{\sqrt{3}}{2}k_y\right) + 1\right]$}. Note that our main results will also be applicable to more generic situations, including honeycomb and Kagome lattices.

From Fig.~\ref{nesting}(b), one sees that each $\bk$ on a FS segment along the $\mathbf{Q}_\alpha$ direction has at least two counterparts $-\bk\pm\mathbf{Q}_\alpha$ to form finite momentum Cooper pairs.
Moreover, $M$ and $X$ points have four momenta for pairing. These mean that the nesting feature allows \emph{full pairing} in the region near the FS, which is in contrast with generic FS without nesting.  

\emph{Time reversal symmetry. ---} The TRS of the Hamiltonian is respected if and only if $\xi_{-\bk}=\xi_{\bk}$ and $\Delta_{{\mathbf{Q}_\alpha}}^{\ast}(\bk+\mathbf{Q}_\alpha)=\Delta_{-\mathbf{Q}_\alpha}(\bk)$.
For the aforementioned form of $\Delta_{\pm \mathbf{Q}_\alpha}(\bk)$, sufficient and necessary conditions for TRS reduce to $\Delta_{{\mathbf{Q}_\alpha}}^{\ast}=\Delta_{{-\mathbf{Q}_\alpha}}$.

The finite momentum pairing 
leads to a spatially varying pairing function 
in real space, resulting in a PDW. To be simple, we set $\Delta_{{\pm \mathbf{Q}_\alpha}}=\Delta e^{i\theta_\alpha} e^{\pm i\frac{\phi_{\alpha}}{2}}$, 
where $\Delta$ is real and positive, $\theta_\alpha \in (-\pi, \pi]$
and $\phi_{\alpha} \in (-\pi, \pi]$.
Thus the pairing function $\Delta(\br)$ reads
\begin{equation}\label{eq:LO}
		\begin{split}
				\Delta(\br) &=2\Delta\sum_{\alpha} e^{i\theta_\alpha}
				\cos \left(
				\mathbf{Q}_{\alpha}\cdot\br + \frac{\phi_{\alpha}}{2}
				\right).
		\end{split}
\end{equation}

\emph{Commensurate PDW and descendant CDW order. ---}
When the PDW melts partially, i.e., the $U(1)$ gauge symmetry is restored but not the translational symmetry, a descendant CDW order will arise with wave vectors $\bq=\pm\mathbf{Q}_\alpha\pm\mathbf{Q}_\beta\neq{}\mathbf{0}$. These 
wave vectors can be classified into three sets: 
(B) $\bq=\pm\mathbf{Q}_{\alpha}$ associated with a $4\times4$ CDW; (C) $\bq=\pm2\mathbf{Q}_{\alpha}$ associated with a $2\times2$ CDW; (D) $\bq=\pm(\mathbf{Q}_{\alpha}-\mathbf{Q}_{\beta})$ associated with a $\frac{4}{\sqrt{3}}\times\frac{4}{\sqrt{3}}$ CDW [see Fig.~\ref{EDOS}(c)].
Note that the FS is nested by 
$\bq=\pm2\mathbf{Q}_{\alpha}$ in C but not those in B or D. So that the  descendant CDW order can be of $4\times{}4$ and $\frac{4}{\sqrt{3}}\times\frac{4}{\sqrt{3}}$, in addition to the $2\times{}2$ CDW 
originating from the FS nesting.

\emph{Quasi-particles. ---}
To study quasi-particle excitations in such a PDW associated with a $4\times{}4$ 
folded BZ, we introduce
	\begin{equation*}
		\begin{array}{ll}
			\hat{C}^{\dagger}_{\bk,\sigma}=&\Big(
			c^{\dagger}_{\bk,\sigma},
			c^{\dagger}_{\bk+\mathbf{Q}_1,\sigma},c^{\dagger}_{\bk-\mathbf{Q}_1,\sigma},
			c^{\dagger}_{\bk+\mathbf{Q}_2,\sigma},c^{\dagger}_{\bk-\mathbf{Q}_2,\sigma},c^{\dagger}_{\bk+\mathbf{Q}_3,\sigma},c^{\dagger}_{\bk-\mathbf{Q}_3,\sigma},
			\\
			&
			c^{\dagger}_{\bk+2\mathbf{Q}_1,\sigma},c^{\dagger}_{\bk+2\mathbf{Q}_2,\sigma},c^{\dagger}_{\bk+2\mathbf{Q}_3,\sigma},c^{\dagger}_{\bk+\mathbf{Q}_1-\mathbf{Q}_2,\sigma},c^{\dagger}_{\bk-\mathbf{Q}_1+\mathbf{Q}_2,\sigma},
			\\
			&	c^{\dagger}_{\bk+\mathbf{Q}_2-\mathbf{Q}_3,\sigma},c^{\dagger}_{\bk-\mathbf{Q}_2+\mathbf{Q}_3,\sigma},c^{\dagger}_{\bk+\mathbf{Q}_3-\mathbf{Q}_1,\sigma},c^{\dagger}_{\bk-\mathbf{Q}_3+\mathbf{Q}_1,\sigma}
			\Big),
		\end{array}
	\end{equation*}
and rewrite Eq.~\eqref{H} in a matrix form:
\begin{subequations}
	\begin{align}
		\begin{split}
			H =&\frac{1}{16}\sum_{\bk}H_{\bk} + \sum_{\bk} \xi_\bk\\
			=&\frac{1}{16}\sum_{\bk}\left(\hat{C}^{\dagger}_{\bk,\uparrow},\hat{C}_{-\bk,\downarrow}\right)\hat{\ham}_{\bk}
			\left(
			\begin{matrix}
				\hat{C}_{\bk,\uparrow}  \\
				\hat{C}^{\dagger}_{-\bk,\downarrow}
			\end{matrix}
			\right) + \sum_{\bk} \xi_\bk,
		\end{split}\\
		\hat{\ham}_{\bk}=&
		\left(
		\begin{matrix}
			\hat{D}(\mathbf{k})  &\hat{\Delta}(\bk)\\
			\hat{\Delta}^{\dagger}(\bk) &-\hat{D}(-\mathbf{k})
		\end{matrix}
		\right).\label{Hk}
	\end{align}
\end{subequations}
Here $\hat{\ham}_{\bk}$ is a $32 \times 32$ matrix, $\hat{D}(\bk)=\text{diag}\{\xi_{\bk_i}\}$,
$\bk_i$ is the $i$-th momentum in $\hat{C}^{\dagger}_{\bk,\uparrow}$, and $\hat{\Delta}(\bk)$ is a $16 \times 16$ matrix defined by $\Delta_{\pm\mathbf{Q}_\alpha}(\bk)$.

The diagonalization of $\hat{\ham}_{\bk}$ leads to 
	\begin{equation}
	\begin{aligned}
		H_{\bk} =\sum_{i=1}^{16}
		E(\bk)^{+}_{i}
		\gamma^{\dagger}_{\bk,\uparrow,i}\gamma_{\bk,\uparrow,i}
		+E(\bk)^{-}_{i} \left(-\gamma^{\dagger}_{-\bk,\downarrow,i}\gamma_{-\bk,\downarrow,i}
		+1
		\right),
	\end{aligned}
\end{equation}
where $E(\bk)^{+(-)}_{i}\left[i\,(17-i)=1,\cdots,16\right]$ are quasi-particles (holes) energy spectra arranged in ascending order.
The particle-hole symmetry (PHS) is manifested by $E(\bk)^{+}_{i}=-E(-\bk)^{-}_{i}$, which would give rise to $E(\bk)^{+}_{i}=-E(\bk)^{-}_{i}$ if the TRS was respected.
$\gamma_{\bk,\uparrow(\downarrow),i}$'s are Bogoliubov quasi-particle operators, and 
$C^{\dagger}_{\bk,\uparrow,i}$ can be written in terms of them,
\begin{align}\label{Bogoliubov}
C^{\dagger}_{\bk,\uparrow,i}&=\sum_{j=1}^{16}\left(u(\bk)_{ij}\gamma^{\dagger}_{\bk,\uparrow,j}+v(\bk)_{ij}\gamma_{-\bk,\downarrow,j}\right),
\end{align}
where $u(\bk)_{ij}$ and $v(\bk)_{ij}$ form a unitary transformation. It is easy to verify that: $E(\bk)_{i}^{\pm}=E(\bk \pm \mathbf{Q}_{\alpha})_{i}^{\pm}$ and $\gamma_{\bk,\sigma,i}=\gamma_{\bk \pm \mathbf{Q}_\alpha,\sigma,i}$, i.e., the BZ is of $4\times{}4$ folding.

\emph{$\mathbb{Z}_2$ symmetry. ---} It is found that there exists additional $\mathbb{Z}_2$ symmetries associated with a theorem as follows.

 {\bf Theorem:} For each $\alpha$, the transformation $\Delta_{\pm\mathbf{Q}_\alpha}\mapsto-\Delta_{\pm\mathbf{Q}_\alpha}$ does not change the energy spectra of the system.

The proof of the theorem can be found in the Supplementary Material~\cite{appendix}. This theorem suggests a $\mathbb{Z}_2\times \mathbb{Z}_2$ symmetry, since only two of $\mathbf{Q}_\alpha$ are independent.

\begin{figure}[tb]
	\includegraphics[width=\linewidth]{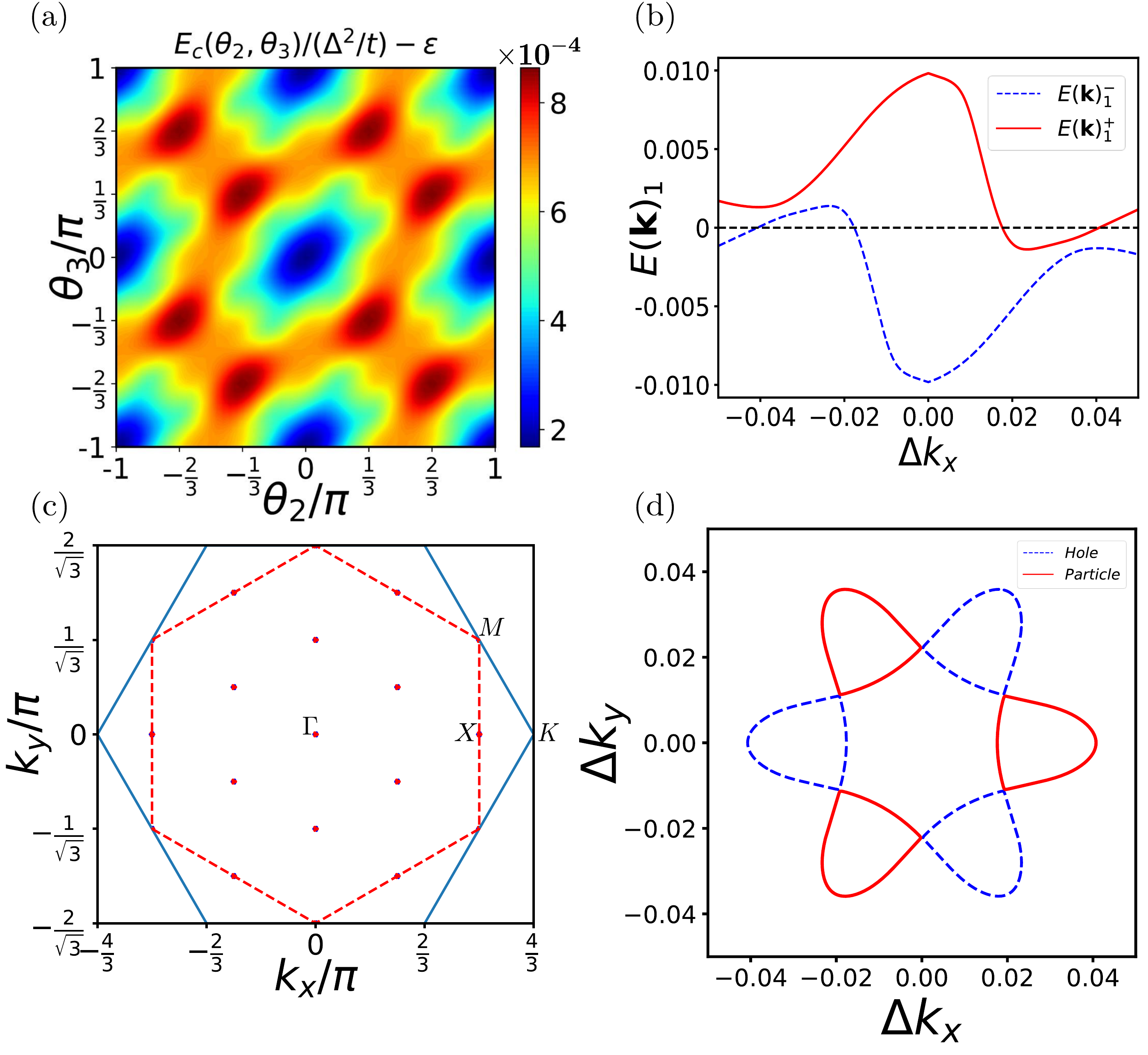}
	\caption{(a) Condensation energy $E_c$ (with an offset $\varepsilon = 0.993$) as a function of $\theta_2$ and $\theta_3$, where $\theta_1=0$ and $\phi_{\alpha}=\pi/2$ have been set [see Eq.~\eqref{eq:LO}]. 
Maximal $E_c$ occurs at $\theta_2=\theta_3-\theta_2\equiv\pm{}2\pi/3\,(\mathrm{mod}\,\pi)$. (b)-(d) Energy dispersion and Bogoliubov Fermi pockets for the lowest energy state: $\phi_{\alpha}=\pi/2$, $\theta_1=0$, $\theta_2=2\pi/3$ and $\theta_3=-2\pi/3$. (b) 
 $E(\bk)^{\pm}_{1}$ around $X$ point that are plotted along 
$\Gamma-X-K$. (c) Bogoliubov Fermi pockets at $M$ and $X$ points and their periodic replica due to the PDW. (d) Quasi-particle (hole)  pocket around $X$ point.
}\label{fig:pocket}
\end{figure}

\emph{Approximate $E(\bk)^{+}_1$. ---} To get insight into low energy excitations, we 
inspect the lowest branch of quasi-particle dispersion, $E(\bk)^{+}_1$, along the FS. Without loss of generality, we consider the FS segment $M-X$ [see Figs.~\ref{nesting}(a) and \ref{nesting}(b)].
For each $\bk$ on this segment and away from $M$ and $X$ points, 
it is found that only the pairings between $\bk$ and $-\bk\pm\mathbf{Q}_1$ are of order of $\Delta$, while other pairing terms are much smaller than them because of the energy cutoff. Keeping 
sizable terms in $\hat{\Delta}(\bk)$ and neglecting others, we find that $\hat{\ham}_\bk$ can be approximately decomposed into 
paired and unpaired parts, i.e., $\hat{\ham}_\bk\approx\hat{\ham}_\bk^{p}\oplus\hat{\ham}_\bk^{f}$~\cite{appendix}. Therefore $E(\bk)^{+}_1$ can be estimated as
\begin{subequations}\label{eq:Ek1}
\begin{equation}
E(\bk)^{+}_1 \approx \text{min}\{E^{p}(\bk), E^{f}(\bk)\},
\end{equation}
where $E^{p(f)}(\bk)$ is the lowest non-negative eigenvalue of $\hat{\ham}_\bk^{p(f)}$.
Straightforward algebra~\cite{appendix} leads to
\begin{equation}\label{Ep}
	E^{p}(\bk) = 2 \Delta\text{min}\left\{ \left|\sin\left(\frac{\phi_{1}}{2}\right)\right|,\left|\cos\left(\frac{\phi_{1}}{2}\right)\right|\right\}.
\end{equation}
\end{subequations}
It takes the minimum $E^{p}(\bk)_\text{min} = 0$ at $\phi_{1}=0,\pi$ and the maximum $E^{p}(\bk)_\text{max} = \sqrt{2} \Delta$ at $\phi_{1}=\pm \frac{\pi}{2}$. Thus, 
the condensation of Cooper pairs will gain most energy at $\phi_{\alpha}=\pm \frac{\pi}{2}$.
Meanwhile, $E^{f}(\bk)$ 
determined by $\hat{\ham}_\bk^{f}$
is responsible for (nearly) unpaired electrons and in-gap excitations in $E(\bk)^{+}_1$ as long as $E^{f}(\bk) < E^{p}(\bk)$. The combination of $E^{p}(\bk)$ and $E^{f}(\bk)$ gives rise to 
$E(\bk)^{+}_1$ approximately.

Away from the FS or near $M$ or $X$ point, other pairing terms become considerable and the simple decomposition of $\hat{\ham}_{\bk}$ does not work any more. 
We shall diagonalize $\hat{\ham}_\bk$ numerically, and study the ground state 
and low energy excitations. 
Hereafter we set $\Lambda=0.1$ and $\Delta=0.02$, unless otherwise specified.

{\em Condensation energy. ---}
The condensation energy $E_c\equiv{}E_n-E_s$, that defined by the energy difference between a SC ground state and corresponding normal state~\cite{SchriefferBook}, has been found as $E_c=\frac{1}{N}\sum_{\bk}\left(\frac{1}{16}\sum_{i=1}^{16}\sum_{s=\pm}\sum_{E(\bk)^{s}_{i}>0}E(\bk)^{s}_{i}-|\xi_{\bk}|\right)$.
The numerical calculation finds that $E_c[\theta_1,\theta_2,\theta_3,\phi_{\alpha}]$ reaches local maxima at $\phi_{\alpha}=\pm\pi/2$. This agrees with the above analysis of approximate $E(\bk)^{+}_1$ [see Eq.~\eqref{eq:Ek1}] well.
Moreover, as shown in Fig.~\ref{fig:pocket}~(a), the PDW state acquires maximum $E_c$ at $\phi_{\alpha}=\pm\pi/2$ and $\theta_2-\theta_1=\theta_3-\theta_2\equiv \pm2\pi/3\,(\mathrm{mod}\,\pi)$, breaking the TRS spontaneously. Owing to the $\mathbb{Z}_2$ symmetry theorem, ($\theta_\alpha\mapsto\theta_\alpha\pm\pi$), the period is $\pi$ instead of $2\pi$ here.

\emph{Ginzburg-Landau free energy.---} The TRS breaking and the $\mathbb{Z}_2$ symmetry 
can be verified in Ginzburg-Landau (GL) theory. Up to quartic order in $\Delta_{\mathbf{Q}_\alpha}$, the GL free energy can be written as~\cite{appendix},
\begin{equation}\label{eq:GL}
\mathcal{F}[\Delta_{\mathbf{Q}_\alpha}] = \mathcal{F}^{(0)} + \mathcal{F}^{(2)}[\Delta_{\mathbf{Q}_\alpha}] + \mathcal{F}^{(4)}[\Delta_{\mathbf{Q}_\alpha}],
\end{equation}
where $\mathcal{F}^{(0)}$ is $\Delta_{\mathbf{Q}_\alpha}$-independent,
$\mathcal{F}^{(2)}=g^{(2)}\sum_{\alpha=1}^{3}
\left(
\left|\Delta_{\mathbf{Q}_{\alpha}} \right|^{2}+\left|\Delta_{-\mathbf{Q}_{\alpha}} \right|^{2}
\right)$ 
with $g^{(2)}<0$,
and 
$\mathcal{F}^{(4)} = \mathcal{F}^{(4)}_{0} + \mathcal{F}^{(4)}_{\phi} + \mathcal{F}^{(4)}_{\theta}$. Here $\mathcal{F}^{(4)}_{0}$ depends on
$|\Delta_{\mathbf{Q}_\alpha}|$ only. $\mathcal{F}^{(4)}_{\phi}$ and $\mathcal{F}^{(4)}_{\theta}$ 
read
\begin{subequations}\label{eq:F4}
\begin{eqnarray}
\mathcal{F}^{(4)}_{\phi} & = & g^{(4)}_{\phi}
    \left[
\left(\Delta^{2}_{\mathbf{Q}_1}\right) \left(\Delta^{2}_{-\mathbf{Q}_1}
\right)^{*}
+\left(\Delta^{2}_{\mathbf{Q}_2}\right) \left(\Delta^{2}_{-\mathbf{Q}_2}
\right)^{*}\right.\nonumber\\
&& \left.+\left(\Delta^{2}_{\mathbf{Q}_3}\right) \left(\Delta^{2}_{-\mathbf{Q}_3}
\right)^{*}+c.c.
    \right]
\end{eqnarray}
and
\begin{eqnarray}
\mathcal{F}^{(4)}_{\theta} & = & g^{(4)}_{\theta}
    \left[
    \left(\Delta_{\mathbf{Q}_1}\Delta_{-\mathbf{Q}_1}
    \right)
    \left(\Delta_{\mathbf{Q}_2}\Delta_{-\mathbf{Q}_2}
    \right)^{*}
    +\left(\Delta_{\mathbf{Q}_2}\Delta_{-\mathbf{Q}_2}
    \right)
    \left(\Delta_{\mathbf{Q}_3}\Delta_{-\mathbf{Q}_3}
    \right)^{*}\right.\nonumber\\
    &  &  +\left.\left(\Delta_{\mathbf{Q}_3}\Delta_{-\mathbf{Q}_3}
    \right)
    \left(\Delta_{\mathbf{Q}_1}\Delta_{-\mathbf{Q}_1}
    \right)^{*}+c.c.
    \right],
\end{eqnarray}
\end{subequations}
respectively, where both $g^{(4)}_{\phi}$ and $g^{(4)}_{\theta}$ are found to be positive and $\Delta_{\mathbf{Q}_\alpha}$-independent~\cite{appendix}. Putting $\Delta_{{\pm \mathbf{Q}_\alpha}}=\Delta e^{i\theta_\alpha} e^{\pm i\frac{\phi_{\alpha}}{2}}$ into the above leads to $\mathcal{F}^{(4)}_{\phi}=2g^{(4)}_{\phi}\Delta^{4}
\sum_{\alpha=1}^{3}\cos \left(2\phi_{\alpha}\right)$ and $\mathcal{F}^{(4)}_{\theta}=2g^{(4)}_{\theta}\Delta^{4}\left[
		\cos \left(2\theta_2-2\theta_1\right)+\cos \left(2\theta_3-2\theta_2\right)+\cos \left(2\theta_1-2\theta_3\right)
		\right]$. Thus, the lowest free energy is achieved at $\phi_{\alpha}=\pm\pi/2$ and  $\theta_2-\theta_1=\theta_3-\theta_2\equiv \pm2\pi/3\,(\mathrm{mod}\,\pi)$.

Henceforward, we shall focus on the lowest energy state with $\theta_1=0$, $\theta_2=2\pi/3$, $\theta_3=-2\pi/3$ and $\phi_{\alpha}=\pi/2$, and study various electronic spectra. 

\emph{Bogoliubov Fermi pockets. ---} As shown in Fig.~\ref{fig:pocket}~(b), around $M$ and $X$ points, 
$E(\bk)^{+}_{1}$ sinks down while 
$E(\bk)^{-}_{1}$ rises up, such that both of them go across  zero energy. This means that 
quasi-particles (holes) possess FS indeed, 
namely, Bogoliubov Fermi 
pockets come into being. These Fermi pockets are located at $M$ and $X$ points and their periodic replica by the PDW (shifted by 
$\bq=\pm\mathbf{Q}_\alpha\pm\mathbf{Q}_\beta\neq{}\mathbf{0}$), as indicated in Fig.~\ref{fig:pocket}~(c). It is displayed in Fig.~\ref{fig:pocket}~(d) that these Fermi pockets exhibit $D_3$ symmetry.

\begin{figure}[tb]
	\includegraphics[width=\linewidth]{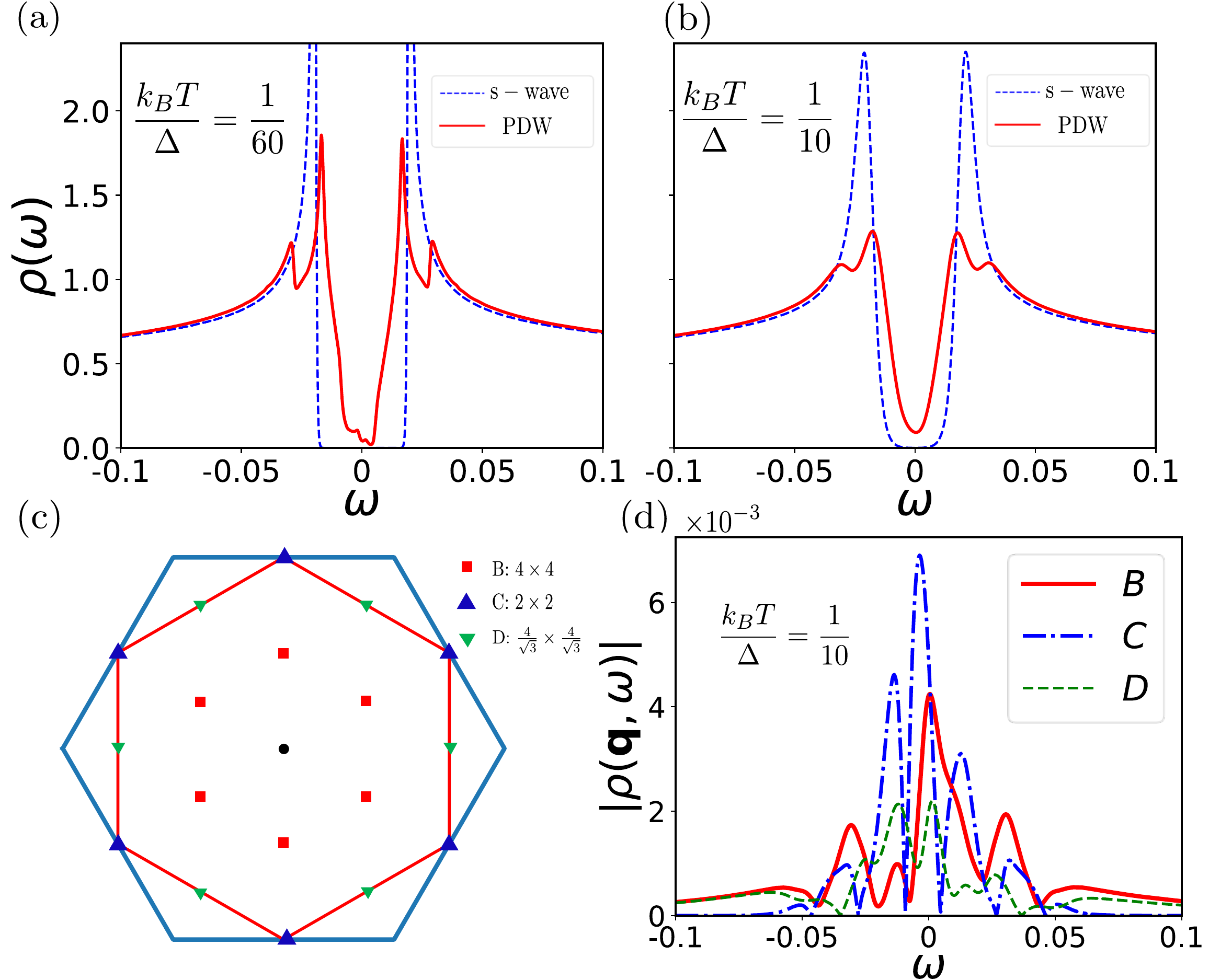}
	\caption{DOS and LDOS. DOS $\rho(\omega)$ at (a) $k_B{}T/\Delta=1/60$ and (b) $k_B{}T/\Delta=1/10$. (c) Wave vectors $\bq=\pm\mathbf{Q}_\alpha\pm\mathbf{Q}_\beta$ (or their equivalent vectors in first BZ) associated with descendant CDW orders.
(d) LDOS $\rho(\bq,\omega)$ exhibits three types of CDW orders B, C and D. Here $\Lambda=0.1$ and $\Delta=0.02$ have been chosen.}\label{EDOS}
\end{figure}
{\em Density of States. --- }
The differential conductance $dI/dV$ measured by STM~\cite{STM_RMP} is proportional to the DOS
that reads 
$\rho(\omega)=-\frac{1}{8N}\sum_{\bk}\sum_{i,j=1}^{16} \Big[|u(\bk)_{ij}|^2\frac{\partial n_{F}\left(\omega-E(\bk)_j^{+}\right)}{\partial \omega} +|v(\bk)_{ij}|^2\frac{\partial n_{F}\left(\omega-E(\bk)_j^{-}\right)}{\partial \omega}\Big]$,
where $n_{F}$ 
is the Fermi distribution, and $u(\bk)_{ij}$ and $v(\bk)_{ij}$ are found via Eq.~\eqref{Bogoliubov}.
As demonstrated in Fig.~\ref{EDOS}, for $k_B{}T=\Delta/60\ll\Delta$, $\rho(\omega)$ exhibits a mini-gap inside the SC gap manifested by sharp coherence peaks; while for $k_B{}T=\Delta/10\lesssim\Delta$, $\rho(\omega)$ (thereby $dI/dV$) curve is of V-shape. 
Note that both the mini-gap and the V-shape DOS suggest electronic excitations inside the SC gap. The extra peaks outside sharp coherence peaks are attributed to the Van Hove singularity~\cite{Van_Hove}, and the asymmetry between $\rho(\omega)$ and $\rho(-\omega)$ is due to the broken particle-hole symmetry in $\xi_{\bk}$.

\emph{Local density of states. ---}
Now we study the LDOS 
that serves as a standard tool to identify CDW orders by STM. 
The Fourier transformation of LDOS for the PDW state is given by $\rho(\bq,\omega)= -\frac{1}{8N}\sum_{\bk}\sum_{i,j=1}^{16} \Big[
u(\bk)_{ij}u^*(\bk+\bq)_{ij}\frac{\partial n_{F}\left(\omega-E(\bk)_j^{+}\right)}{\partial \omega}+ v(\bk)_{ij}v^*(\bk+\bq)_{ij}\frac{\partial n_{F}\left(\omega-E(\bk)_j^{-}\right)}{\partial \omega} \Big] \bar{\delta}_{\bk, \bk + \bq}$, where 
$\bar{\delta}_{\bk,\bk^{\prime}}\equiv\sum_{n,m=-\infty}^{\infty}\delta_{\bk+n\mathbf{Q}_1+m\mathbf{Q}_2,\bk^{\prime}}$~\cite{appendix}. It has been found that $\rho(\bq,\omega)$ does not vanish only at finite number of $\bq$-points in BZ, as labeled in Fig.~\ref{EDOS}(c). These $\bq$-points are nothing but 
wave vectors of the descendant CDW order.
Note that $|\rho(\bq,\omega)|$ takes the same value at $\bq$-points within each set of B, C or D~\cite{appendix}. As demonstrated in Fig.~\ref{EDOS}(d), $|\rho(\bq,\omega)|$ displays both (B) $4\times4$ and (D) $4/\sqrt{3}\times4/\sqrt{3}$ CDW orders in addition to (C) $2\times2$ CDW order caused by the FS nesting.
Define the integrated intensity of these CDW orders as $I_{CDW}(\bq)=\left|\int{}\rho(\bq,\omega)d\omega\right|/\left|\int\rho(\bq=\mathbf{0},\omega)d\omega\right|$, we find that $I_{CDW}^{B,C,D}=1.51\times10^{-5}, 1.02\times10^{-5}, 2.82\times10^{-6}$
at $k_B{}T=\Delta/10$ for the three types of CDW orders respectively.

\begin{figure}[tb]
	\includegraphics[width=\linewidth]{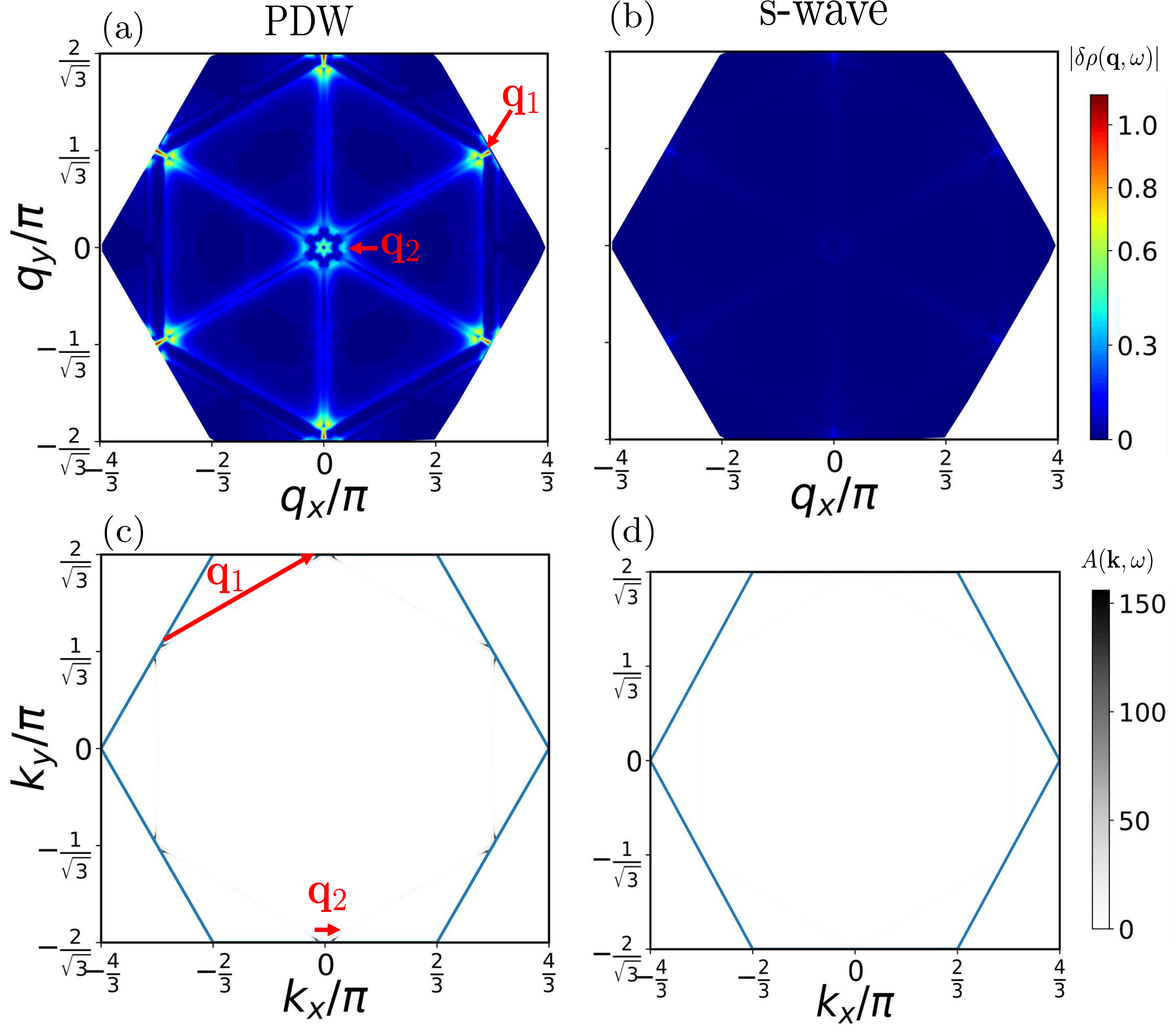}
	\caption{LDOS modulation $|\delta \rho(\bq, \omega)|$ for (a) the PDW state and (b) a uniform s-wave SC state. Electron spectral function $A(\bk, \omega)$ for (c) the PDW state and (d) a uniform s-wave state. Wave vectors $\bq_1$ and $\bq_2$ in (a) and (c) indicate dominant scattering processes connecting two points with large $A(\bk, \omega)$ (and their symmetric equivalence). Here $\omega=0.01<\Delta=0.02$ has been chosen.}\label{QPI}
\end{figure}

{\em Quasi-particle interference. --- }
In the presence of elastic scatterings, the LDOS will be modulated due to the effect of QPI. To characterize this feature, we follow Ref.~\cite{Wang_QPI} to study the modulated LDOS $\delta \rho (\br, \omega)$, or its Fourier transformation 
that is given by
\begin{equation}\label{mLDOS}
	\begin{split}
		& \delta \rho (\bq, \omega) \equiv \rho_s (\bq, \omega) - \rho (\bq, \omega)\\
		= - &\frac{1}{16\pi{}N}\sum_{\bk}\text{Im}\widetilde{\text{Tr}}
		\left[
		\hat{\mathcal{G}}(\bk+\bq,\omega + i \delta)\hat{T}(\omega)
		\hat{\mathcal{G}}(\bk,\omega + i \delta)
		\right],
	\end{split}
\end{equation}
where $\rho_s$ ($\rho$) 
is the LDOS 
in the presence (absence) of scatterings.
$\widetilde{\text{Tr}}$ means tracing the upper-left $16 \times 16$ block in a $32\times{}32$ matrix.
$\hat{\mathcal{G}}(\bk,\omega + i \delta)=\left[(\omega + i \delta)\mathcal{I} - \hat{\ham}_{\bk}\right]^{-1}$ is Green's function in the absence of scatterings and $\hat{T}(\omega)=\left[(V_s\hat{\tau}_3)^{-1} - \frac{1}{N}\sum_{\bk}\hat{\mathcal{G}}(\bk,\omega + i \delta)\right]^{-1}$ is the scattering matrix.
Here $V_s$ is the nonmagnetic scattering impurity strength, and $\hat{\tau}_3$ is the Pauli matrix spanning Nambu space.

The modulation $|\delta \rho (\bq, \omega)|$ with $V_s = 0.1$ at $\omega=0.01(<\Delta=0.02)$ is plotted in Fig.~\ref{QPI}(a). For comparison, we also study QPI of a uniform s-wave superconductor, 
as shown in Fig.~\ref{QPI}(b). In both figures the intensity at $\bq = \mathbf{0}$ has been subtracted.

\emph{Electron spectral function. ---}
The LDOS modulation due to scatterings can be analyzed by electron spectral function $A(\bk, \omega) = -\frac{1}{\pi}\text{Im}[\hat{\mathcal{G}}(\bk, \omega + i \delta)]_{11}$
in the absence of scattering.
As is pointed out in Ref.~\cite{Wang_QPI}, the 
summation in Eq.~\eqref{mLDOS} is dominated by terms in which both $\bk$ and $\bk + \bq$ are poles of $\hat{\mathcal{G}}$. Thus the vectors $\bq$ associated with the scattering processes connecting two points with large $A(\bk, \omega)$ will show significant $|\delta \rho (\bq, \omega)|$. This feature of $\bq$ is displayed in Fig.~\ref{QPI}(c). An essential difference between the PDW state and a uniform s-wave state is that in-gap state is absent in the latter and the corresponding $A(\bk,\omega)$ and $|\delta \rho (\bq, \omega)|$ 
vanish at $\omega<\Delta$, as shown in Fig.~\ref{QPI}(b) and (d). This also provides an experiment scheme to probe PDW states.

\emph{Discussions and conclusions. ---}

(i) Recently discovered Kagome SC AV$_3$Sb$_5$ (A=K,Rb,Cs) with a nearly $3/4$ filled electron band is a natural platform towards the realization of the interplay between PDW and FS nesting~\cite{CVS,KVS,RVS,jiang2021kagome,prx-Stephen,apl-Michelle}. 
TRS breaking signatures have been extensively discussed both experimentally and theoretically in AV$_3$Sb$_5$~\cite{hasan,yu2021evidence,mielke2021timereversal,fengxl,rahul,balents}. 
For the SC properties, the AV$_3$Sb$_5$ is shown to be a spin-singlet SC  hosting s-wave features~\cite{nmr,huiqiu_yuan,donglai_feng}. However, a residual thermal transport at $T=0$ and ``multi-gap''  V-shape DOS with residual zero-energy contributions were observed 
in SC states~\cite{shiyan_li,donglai_feng,hongjun_gao,zhenyu_wang}, which conflicts with the conventional s-wave nature. This contradiction can be resolved within the TRS breaking PDW scenario proposed in the present work. More importantly, a PDW state ordering at $\mathbf{Q}_\alpha$ has been observed in recent 
STM measurements~\cite{hongjun_gao}. Therefore, our theory may provide new insight into  the PDW states and TRS breaking in AV$_3$Sb$_5$. Indeed, both $2\times2$ and $4\times4$ CDW have been observed in STM. Our theory suggests that the $\frac{4}{\sqrt{3}}\times\frac{4}{\sqrt{3}}$ CDW should appear as well, as long as the frequency $\omega$ is chosen properly.

(ii) Indeed, such a TRS breaking SC state breaks the spatial inversion symmetry as well [see Eq.~\eqref{eq:LO}], resulting in a chiral state with stable residual gapless 
quasi-particle excitations. 
The ground state is a flux state with spontaneous loop current~\cite{flux,PDW-loop-current}, as calculated in the Supplementary Material~\cite{appendix}. And the Bogoliubov Fermi pockets yield the linear $T$-dependent specific heat at low temperature.   

(iii) One of remaining issues is what microscopic theory may give rise to the finite-momentum Cooper pairing instability 
on a nested FS. In the weak interaction limit, pairing at zero momentum is usually favored. Nonetheless, strong correlation might favor PDW instability against uniform pairing (see, e.g. Ref.~\cite{ZYHan-PRL2020, t-J-U}). By establishing the microscopic model, the comparison with relevant models~\cite{Honerkamp03,Wang13,Nandkishore14,CDW-re-SC} based on the conventional CDW instabilities with nesting vector $2\mathbf{Q}_{\alpha}$ is one of the essential topics.

In summary,  we have found that the FS nesting allows a full PDW pairing and 
in-gap states simultaneously. Such a PDW ansatz will give rise to a TRS breaking ground state.
Subsequently, descendant CDW orders and various electronic spectra
have been studied, and the relevance to newly discovered Kagome SC 
has been revealed.

\emph{Acknowledgment.---} We thank Hui Chen, Shiyan Li, Zheng Li, Tao Wu, Fu-Chun Zhang, and Tong Zhang for helpful discussions. This work is partially supported by National Natural Science Foundation of China (No. 12274441, No. 12034004, No. 12174428 and No. 11825404), the K. C. Wong Education Foundation (Grant No. GJTD-2020-01), and the Strategic Priority Research Program of Chinese Academy of Sciences (No. XDB28000000).

\bibliography{nestingPDW}

\clearpage


\setcounter{table}{0}
\setcounter{figure}{0}
\setcounter{equation}{0}
\setcounter{page}{1}
\renewcommand{\thetable}{S\arabic{table}}
\renewcommand{\thefigure}{S\arabic{figure}}
\renewcommand{\theequation}{S\arabic{equation}}

\renewcommand{\thesection}{\Roman{section}}
\renewcommand{\thesubsection}{\Alph{subsection}}

\begin{widetext}
	
	\begin{center}
		\textbf{Supplementary Material for "Interplay between Pair Density Wave and a Nested Fermi Surface"}
	\end{center}	
	
	
	This supplementary Material provides more details on our hexagonal lattice model, including the proof of $\mathbb{Z}_2$  symmetry theorem, the calculation of $E(\bk)^{+}_1$ along the FS segment, more numerical study on the condensation energy, ground state degeneracy and the derivation of Ginzburg-Landau free energy  and local density of states.

\section{\uppercase\expandafter{\romannumeral1}. The proof of $\mathbb{Z}_2$  symmetry theorem}

Here we provide more details for the proof of the $\mathbb{Z}_2$ theorem presented in the main text.

 {\bf Theorem:} For each $\alpha$, the transformation $\Delta_{\pm\mathbf{Q}_\alpha}\mapsto-\Delta_{\pm\mathbf{Q}_\alpha}$ does not change the energy spectra of the system.
 	
	\begin{figure}[h]
	\includegraphics[width=8.4cm]{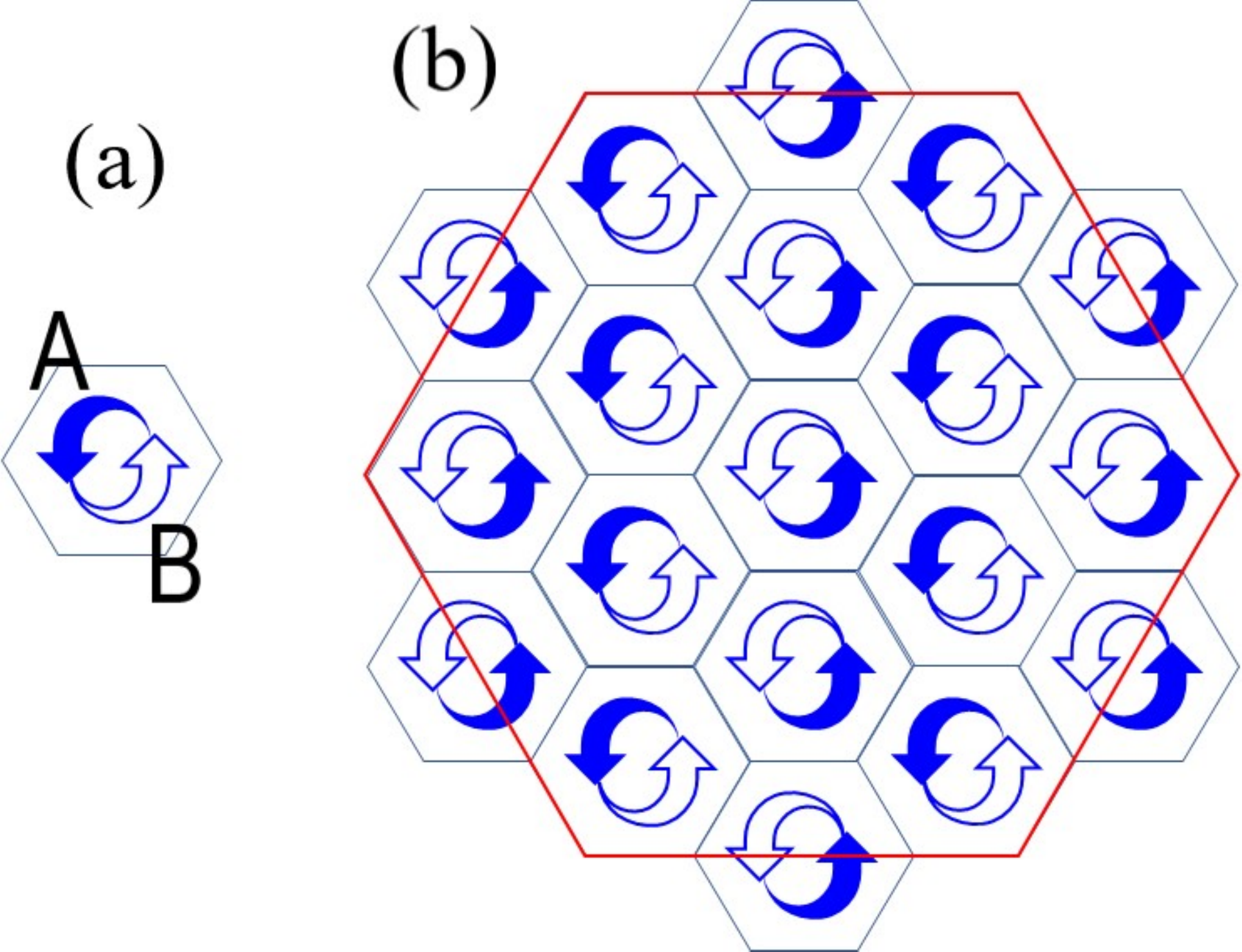}
	\caption{(a) A centrosymmetric bipartition for the $4\times4$ folded BZ. The folded zone is divided into two parts A and B, such that A is the inversion of B. (b) A $\mathbf{Q}_1$ stripy tiling of the reciprocal plane with folded zone and its inversion. They are aligned alternatively along $\mathbf{Q}_2$ and $\mathbf{Q}_3$ directions, while keep $\mathbf{Q}_1$ translational invariant. This tiling gives rise to a centrosymmetric bipartition for the whole BZ, such that $\bk$ and $\mathbf{Q}_1-\bk$ belong to opposite parts (A and B), while $\mathbf{Q}_2-\bk$ and $\mathbf{Q}_3-\bk$ belong to the same part. The red hexagon encloses first (unfolded) BZ.}\label{fig:Z2Q1}
\end{figure}

	{\bf Proof:} Wihout loss of generality, we consider $\alpha=1$ and a centrosymmetric bipartition (A and B) of first BZ as illustrated in Fig.~\ref{fig:Z2Q1}. Define a function $\eta(\bk):=+(-)1$ for $\bk\in\text{A(B)}$, we have $\eta(\mathbf{Q}_2-\bk)=\eta(\mathbf{Q}_3-\bk)=-\eta(\mathbf{Q}_1-\bk)=\eta(\bk)$. Thus, the unitary transformation: $c_{\bk,\uparrow}\mapsto{}c_{\bk,\uparrow}$ and $c_{\bk,\downarrow}\mapsto{}\eta({\bk})c_{\bk,\downarrow}$ gives rise to $\Delta_{\pm\mathbf{Q}_1}\mapsto-\Delta_{\pm\mathbf{Q}_1}$ and $\Delta_{\pm\mathbf{Q}_{2,3}}\mapsto\Delta_{\pm\mathbf{Q}_{2,3}}$, and will not change energy spectra. QED.
	
	\section{\uppercase\expandafter{\romannumeral2}. $E(\bk)^{+}_1$ along the hexagonal fermi surface }\label{ssec:dispersion along fermi surface}
	
	We now study the approximate $E(\bk)^{+}_1$ along the FS. By symmetry, we consider the FS segment $M-X$ [see Figs.~1(a) and 1(b) in the main text] only. For $\bk$ satisfies $k_x=\pi$ and $k_y \in \left(0,\frac{\pi}{\sqrt{3}}\right)$, with sufficiently small energy cutoff $\Lambda$, the decomposition of $\hat{\ham}_\bk$ in Eq.~(3b) is of the following form,
	\begin{subequations}
		\begin{equation}
			H_\bk
			= \left(
			{\hat{C}}^{\dagger p}_{\bk, \uparrow}, {\hat{C}^{p}}_{-\bk,\downarrow}\right)
			\hat{\ham}^{p}_{\bk}
			\left(
			\begin{matrix}
				{\hat{C}^{p}}_{\bk,\uparrow}  \\
				{\hat{C}}^{\dagger p}_{-\bk,\downarrow}
			\end{matrix}
			\right)+\left(
			{\hat{C}}^{\dagger f}_{\bk, \uparrow}, {\hat{C}^{f}}_{-\bk,\downarrow}\right)
			\hat{\ham}^{f}_{\bk}
			\left(
			\begin{matrix}
				{\hat{C}^{f}}_{\bk,\uparrow}  \\
				{\hat{C}}^{\dagger f}_{-\bk,\downarrow}
			\end{matrix}
			\right),
		\end{equation}
		where the pairing part $\hat{\ham}^{p}_{\bk}$ can be further decomposed as
		\begin{equation*}
			\left(
			{\hat{C}}^{\dagger p}_{\bk, \uparrow}, {\hat{C}^{p}}_{-\bk,\downarrow}\right)
			\hat{\ham}^{p}_{\bk}
			\left(
			\begin{matrix}
				{\hat{C}^{p}}_{\bk,\uparrow}  \\
				{\hat{C}}^{\dagger p}_{-\bk,\downarrow}
			\end{matrix}
			\right)=\left(
			c^{\dagger}_{\bk,\uparrow},c^{\dagger}_{\bk+2\mathbf{Q}_1,\uparrow},c_{-\bk+\mathbf{Q}_1,\downarrow},c_{-\bk-\mathbf{Q}_1,\downarrow}
			\right)
			\hat{\ham}^{\tilde{p}}_{\bk}
			\left(
			\begin{matrix}
				c_{\bk,\uparrow}  \\
				c_{\bk+2\mathbf{Q}_1,\uparrow}\\
				c^{\dagger}_{-\bk+\mathbf{Q}_1,\downarrow}\\
				c^{\dagger}_{-\bk-\mathbf{Q}_1,\downarrow}
			\end{matrix}
			\right)
			+\left(
			c^{\dagger}_{\bk+\mathbf{Q}_1,\uparrow},c^{\dagger}_{\bk-\mathbf{Q}_1,\uparrow},c_{-\bk,\downarrow},c_{-\bk+2\mathbf{Q}_1,\downarrow}
			\right)
			\hat{\ham}^{\tilde{p}}_{\bk}
			\left(
			\begin{matrix}
				c_{\bk+\mathbf{Q}_1,\uparrow} \\
				c_{\bk-\mathbf{Q}_1,\uparrow}\\
				c^{\dagger}_{-\bk,\downarrow}\\
				c^{\dagger}_{-\bk+2\mathbf{Q}_1,\downarrow}
			\end{matrix}
			\right).
		\end{equation*}
		Here $\hat{\ham}^{\tilde{p}}_{\bk}$ reads
		\begin{equation}\label{seq:tri-Hp}
			\hat{\ham}^{\tilde{p}}_{\bk}
			= 
			\left(
			\begin{matrix}
				0 &0 &\Delta_{\mathbf{Q}_1} &\Delta_{-\mathbf{Q}_1}\\
				0 &0 &\Delta_{-\mathbf{Q}_1} &\Delta_{\mathbf{Q}_1}\\
				\Delta^{*}_{\mathbf{Q}_1} &\Delta^{*}_{-\mathbf{Q}_1} &0 &0\\
				\Delta^{*}_{-\mathbf{Q}_1} &\Delta^{*}_{\mathbf{Q}_1} &0 &0
			\end{matrix}
			\right)
			= \Delta
			\left(
			\begin{matrix}
				0 &0 &e^{i\frac{\phi_{1}}{2} + i \theta_1} &e^{-i\frac{\phi_{1}}{2} + i \theta_1}\\
				0 &0 &e^{-i\frac{\phi_{1}}{2} + i \theta_1} &e^{i\frac{\phi_{1}}{2} + i \theta_1}\\
				e^{-i\frac{\phi_{1}}{2} - i \theta_1} &e^{i\frac{\phi_{1}}{2} - i \theta_1} &0 &0\\
				e^{i\frac{\phi_{1}}{2} - i \theta_1} &e^{-i\frac{\phi_{1}}{2} - i \theta_1} &0 &0
			\end{matrix}
			\right).
		\end{equation}
		${\hat{C}^{\dagger f}}_{\bk,\uparrow}=
		\left(
		c^{\dagger}_{\bk+\mathbf{Q}_2,\uparrow},c^{\dagger}_{\bk-\mathbf{Q}_2,\uparrow},c^{\dagger}_{\bk+\mathbf{Q}_3,\uparrow},c^{\dagger}_{\bk-\mathbf{Q}_3,\uparrow},c^{\dagger}_{\bk+2\mathbf{Q}_2,\uparrow},c^{\dagger}_{\bk+2\mathbf{Q}_3,\uparrow},c^{\dagger}_{\bk+\mathbf{Q}_1-\mathbf{Q}_2,\uparrow},c^{\dagger}_{\bk-\mathbf{Q}_1+\mathbf{Q}_2,\uparrow},	c^{\dagger}_{\bk+\mathbf{Q}_2-\mathbf{Q}_3,\uparrow},c^{\dagger}_{\bk-\mathbf{Q}_2+\mathbf{Q}_3,\uparrow},c^{\dagger}_{\bk+\mathbf{Q}_3-\mathbf{Q}_1,\uparrow}, c^{\dagger}_{\bk-\mathbf{Q}_3+\mathbf{Q}_1,\uparrow}\right)$ and
		\begin{equation} \label{seq:tri-Hf}
			\hat{\ham}^{f}_{\bk}=
			\left(
			\begin{matrix}
				\hat{D}^{f}(\mathbf{k})  &0\\
				0 &-\hat{D}^{f}(-\mathbf{k})
			\end{matrix}
			\right),
		\end{equation}
	\end{subequations}
	where $\hat{D}^f(\bk)=\text{diag}\{\xi_{\bk^f_i}\}$ is a diagonal matrix and $\bk^f_i$ is the $i$-th momentum in $\hat{C}^{\dagger f}_{\bk,\uparrow}$.
	
	We define $E^{p}(\bk)$ and $E^{f}(\bk)$ as the lowest non-negative eigenvalues of $\hat{\ham}_\bk^{p}$ and $\hat{\ham}_\bk^{f}$ respectively. The eigenvalues of  Eq.~\eqref{seq:tri-Hp} are $\pm 2 \Delta \sin \left(\frac{\phi_{1}}{2}\right)$ and $\pm 2 \Delta \cos \left(\frac{\phi_{1}}{2}\right)$. Thus $E^{p}(\bk) = 2 \Delta\text{min}\left\{ \left|\sin\left(\frac{\phi_{1}}{2}\right)\right|,\left|\cos\left(\frac{\phi_{1}}{2}\right)\right|\right\}$. From Eq.~\eqref{seq:tri-Hf}, we obtain that $E^{f}(\bk)=\text{min}\{|\xi_{\bk^f_i}| \}$. Hence we have 
	\begin{equation}
		E(\bk)^{+}_1 \approx \text{min}\left\{E^{p}(\bk), E^{f}(\bk)\right\}=\text{min}\left\{
		2 \Delta\text{min}\left\{ \left|\sin\left(\frac{\phi_{1}}{2}\right)\right|,\left|\cos\left(\frac{\phi_{1}}{2}\right)\right|\right\}, \text{min}\{|\xi_{\bk^f_i}| \}		
		\right\}.
	\end{equation}
 
	Notice that the eigenvalues of  Eq.~\eqref{seq:tri-Hp} and Eq.~\eqref{seq:tri-Hf} are $\theta_\alpha$-independent. To study the effect of TRS breaking, we need to consider pairing along all the three directions of $\mathbf{Q}_\alpha$ numerically. 
	
\section{\uppercase\expandafter{\romannumeral3}. Numerical Study on Condensation energy}\label{ssec:condensation energy}
To show that the aforementioned approximation of $E(\bk)^{+}_1$  is reasonable and $E_c$  acquires local maxima at $\phi_{\alpha}=\pm \pi/2$, we study $E_c$ at different $\phi_{\alpha}$ numerically, keeping 
$\Delta=0.02$ and $\Lambda=0.1$. We fix the values of $\theta_{\alpha}$ randomly and find $\phi_{\alpha}$ that maximize $E_c$. For various random initial variables $\phi_{\alpha}$, the results that maximize the condensation energy fit $\phi_{\alpha}=\pm \pi/2$ well.  

	Here we present a numerical result of a special case: keeping $\theta_{\alpha} = 0,\, 2\pi/3,\, -2\pi/3$ and setting $\phi_1=\phi_2=\phi_3=\phi$. The result of $E_c(\phi)$ is shown in Fig.~\ref{sfig:tri-condensation}. We can see that the maximum of $E_c(\phi)$ is found at $\phi=\pm/2$, which is in good agreement with our analysis of approximate $E(\bk)^{+}_1$. 
	
	To further check our results, we maximize the condensation energy $E_c[\theta_2, \theta_3,\phi_{\alpha}]$ numerically, keeping $\theta_1=0$, $\Delta=0.02$ and $\Lambda=0.1$. For various random initial variables $[\theta_2,\, \theta_3,\, \phi_{\alpha}]$, the results that maximize the condensation energy fit $\phi_{\alpha}=\pm \pi/2$ and $\theta_2-\theta_1=\theta_3-\theta_2\equiv \pm2\pi/3\,(\mathrm{mod}\,\pi)$ well.
	
		\begin{figure}[htb]
	\includegraphics[width=0.6\textwidth]{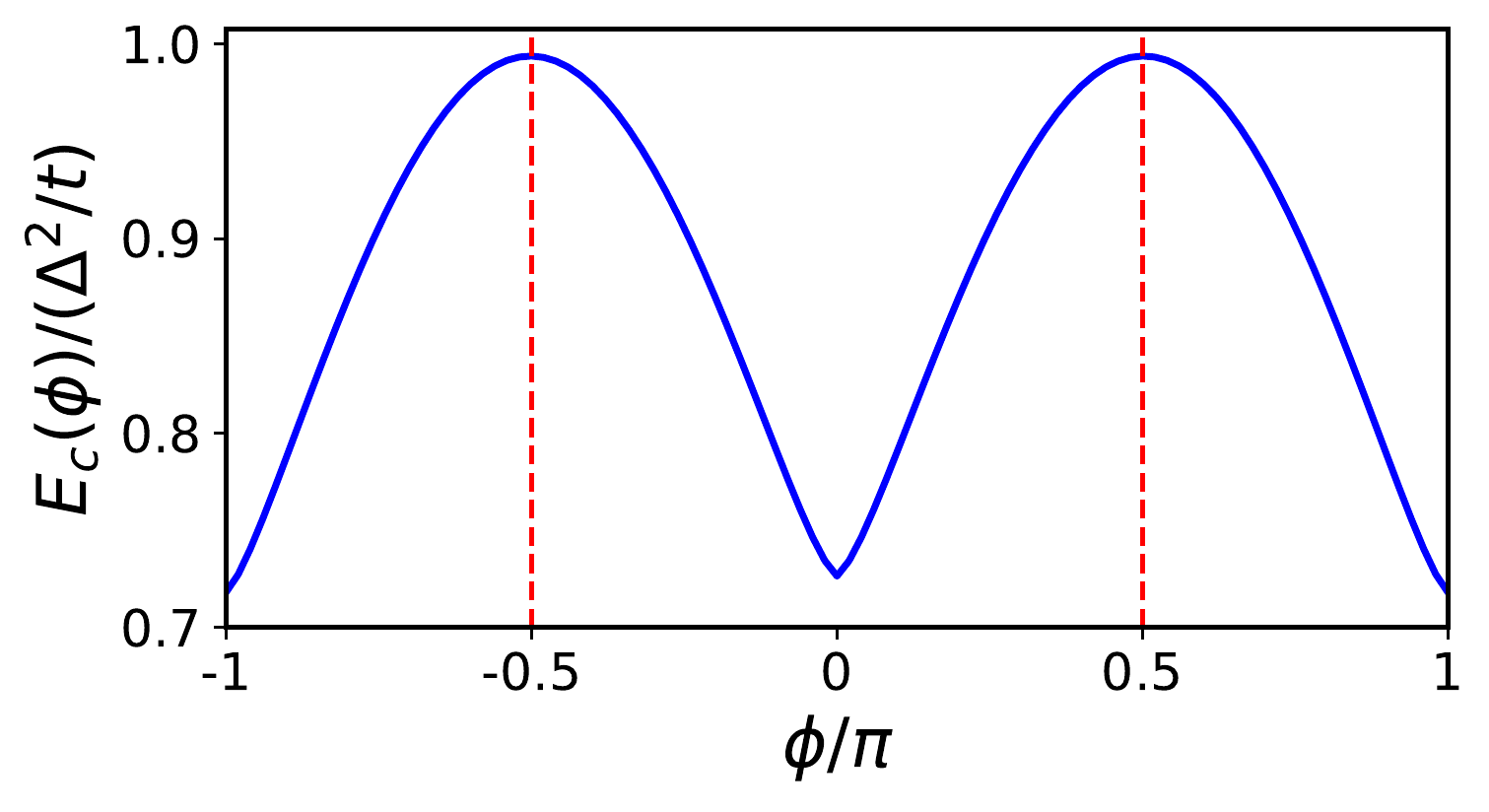}
	\caption{Condensation energy $E_c$ as a function of $\phi$. Here $\theta_1=0,\, \theta_2=2\pi/3,\, \theta_3=-2\pi/3,\,\phi_1=\phi_2=\phi_3=\phi,\,\Delta=0.02$ and $\Lambda=0.1$ have been chosen.}\label{sfig:tri-condensation}
\end{figure}

\section{\uppercase\expandafter{\romannumeral4}. Ground State Degeneracy}\label{ssec:ground-state-degeneracy}
Since the condensation energy is maximized at $\phi_{\alpha}=\pm\pi/2$ and $\theta_2-\theta_1=\theta_3-\theta_2\equiv \pm2\pi/3\,(\mathrm{mod}\,\pi)$, each set of $[\theta_{\alpha},\phi_{\alpha}]$ satisfies these conditions will give rise to a ground state of the system. We now discuss the ground state degeneracy of our PDW state. Notice that a gauge transformation reads $c_{\br,\sigma} \mapsto e^{i\theta_{1}/2} c_{\br,\sigma}$ leads to a transformation of $\theta_{\alpha}$ as follows,
$$[\theta_{1},\theta_{2},\theta_{3}]{\longmapsto}[0, \theta_{2}-\theta_{1},\theta_{3}-\theta_{1}].$$
Thus, we can always set $\theta_{1}=0$ and focus on different sets of $[\theta_{2},\theta_{3},\phi_{\alpha}]$ satisfy $\phi_{\alpha}=\pm\pi/2$ and $\theta_2=\theta_3-\theta_2\equiv \pm2\pi/3\,(\mathrm{mod}\,\pi)$. There are $8$ sets of $[\theta_{2},\theta_{3}]$ satisfy $\theta_2=\theta_3-\theta_2\equiv \pm2\pi/3\,(\mathrm{mod}\,\pi)$ (it can be seen from the $8$ maxima in Fig.~2(a) in the main text) and $8$ sets of $[\phi_{1},\phi_{2},\phi_{3}]$ satisfy $\phi_{\alpha}=\pm\pi/2$. Consequently, the ground state is of $8 \times 8$-fold degeneracy. 

The $64$ different ground states can be related through several symmetry operators. It can be seen from the transformation of $\Delta(\br)$ under the corresponding symmetry operator. Recall the form of $\Delta(\br)$ in Eq.~(2),
$$ 	\Delta(\br)=2\Delta\sum_{\alpha} e^{i\theta_\alpha}
\cos \left(
\mathbf{Q}_{\alpha}\cdot\br + \frac{\phi_{\alpha}}{2}
\right).$$
The  $\mathbb{Z}_2$ transformation $\Delta_{\pm\mathbf{Q}_\alpha}\mapsto-\Delta_{\pm\mathbf{Q}_\alpha}$ gives rise to $\theta_{\alpha} \mapsto \theta_{\alpha} + \pi$ (for a certain $\alpha$). The time reversal transformation leads to $\Delta(\br)\mapsto\Delta^{*}(\br)$ as well as $\theta_{\alpha} \mapsto -\theta_{\alpha}$ (acting on all the three $\alpha$ simultaneously). Thus the $8$ sets of $[\theta_{2},\theta_{3}]$ can be related with each other through these two kinds of transformations. As for the $8$ sets of $[\phi_1,\phi_2,\phi_3]$, we can apply the lattice translation operator: $T_i:\br \mapsto \br + \mathbf{a}_i$ ($\mathbf{Q}_{i} \cdot \mathbf{a}_{j}=\pi/2\delta_{ij},\,i,j=1,2$), the spatial inversion operator $\mathcal{P}:\br \mapsto -\br$ and the $\mathbb{Z}_2$ transformation to change the sign of one $\phi_{\alpha}$ and keep the other two unchanged. This process provides a path links two sets of $[\phi_1,\phi_2,\phi_3]$ and all the $8$ sets of $[\phi_1,\phi_2,\phi_3]$ are related with each other through these paths. As a concrete example, we give the specific process of the transformation $[\pi/2,\pi/2,\pi/2]\mapsto[-\pi/2,\pi/2,\pi/2]$ here. Begin with $[\phi_1,\phi_2,\phi_3]=[\pi/2,\pi/2,\pi/2]$, $\Delta(\br)$ is of the following form
\begin{subequations}
\begin{equation}	
\Delta_{0}(\br)=2\Delta
\left[
\cos \left(
\mathbf{Q}_{1}\cdot\br + \frac{\pi}{4}
\right)
+
e^{i\theta_2}
\cos \left(
\mathbf{Q}_{2}\cdot\br + \frac{\pi}{4}
\right)
+
e^{i\theta_3}
\cos \left(
\mathbf{Q}_{3}\cdot\br + \frac{\pi}{4}
\right)
\right].
\end{equation}
Under $T_2:\br\mapsto \br+\mathbf{a}_2$, $\Delta(\br)$ becomes
\begin{equation}	
	\Delta_{0}(\br)\mapsto\Delta_{1}(\br)=2\Delta
	\left[
	\cos \left(
	\mathbf{Q}_{1}\cdot\br + \frac{\pi}{4}
	\right)
	+
	e^{i\theta_2}
	\cos \left(
	\mathbf{Q}_{2}\cdot\br + \frac{3\pi}{4}
	\right)
	+
	e^{i\theta_3}
	\cos \left(
	\mathbf{Q}_{3}\cdot\br - \frac{\pi}{4}
	\right)
	\right].
\end{equation}
Then the transformation $\Delta_{\pm\mathbf{Q}_2}\mapsto -\Delta_{\pm\mathbf{Q}_2}$ ($\theta_{2}\mapsto\theta_{2}+\pi$) gives rise to 
\begin{equation}	
	\Delta_{1}(\br)\mapsto\Delta_{2}(\br)=2\Delta
	\left[
	\cos \left(
	\mathbf{Q}_{1}\cdot\br + \frac{\pi}{4}
	\right)
	+
	e^{i\theta_2}
	\cos \left(
	\mathbf{Q}_{2}\cdot\br - \frac{\pi}{4}
	\right)
	+
	e^{i\theta_3}
	\cos \left(
	\mathbf{Q}_{3}\cdot\br - \frac{\pi}{4}
	\right)
	\right].
\end{equation}
Finally, we obtain $\Delta(\br)$ with $[\phi_1,\phi_2,\phi_3]=[-\pi/2,\pi/2,\pi/2]$ under the operator $\mathcal{P}$, 
\begin{equation}	
	\Delta_{2}(\br)\mapsto\Delta_{3}(\br)=2\Delta
	\left[
	\cos \left(
	\mathbf{Q}_{1}\cdot\br - \frac{\pi}{4}
	\right)
	+
	e^{i\theta_2}
	\cos \left(
	\mathbf{Q}_{2}\cdot\br + \frac{\pi}{4}
	\right)
	+
	e^{i\theta_3}
	\cos \left(
	\mathbf{Q}_{3}\cdot\br + \frac{\pi}{4}
	\right)
	\right].
\end{equation}
\end{subequations}

\section{\uppercase\expandafter{\romannumeral5}. Ginzburg-Landau free energy}\label{ssec:GL-free-energy}
We begin with the Gorkov Green's function $\mathcal{G}(i \omega_{n}, \bk)$ in our PDW state,
\begin{equation}
	\mathcal{G}^{-1}(i \omega_{n}, \bk) \equiv \mathcal{G}^{-1}_{0}(i \omega_{n}, \bk) +\Sigma (\bk), 
\end{equation}
where
\begin{equation}
	\mathcal{G}^{-1}_{0}(i \omega_{n}, \bk)
	=\left(
	\begin{matrix}
		G^{-1}_{0}(i \omega_n, \bk) & 0\\
		0 & -G^{-1}_{0}(-i \omega_n, -\bk)
	\end{matrix}
	\right), \,
	 \Sigma (\bk)=
	 \left(
	 \begin{matrix}
	 	0 &-\hat{\Delta}(\bk)\\
	 	-\hat{\Delta}^{\dagger}(\bk) &0
	 \end{matrix}
	 \right).
\end{equation}
Here $G_{0}(i \omega_n, \bk)$ is the normal state Green's function. In our PDW state, $G_{0}(i \omega_n, \bk)$ is a $16 \times 16$ matrix satisfies $G_0(i \omega_n, \bk)_{ij}=(i \omega_n - \xi_{\bk_i})^{-1} \delta_{ij}$ ($\bk_{i} $ is the $i$-th momentum in $\hat{C}^{\dagger}_{\bk,\uparrow}$ in the main text). $\hat{\Delta}(\bk)$ is a $16 \times 16$ pairing matrix defined by $\Delta_{\pm \mathbf{Q}_{\alpha}}(\bk)$.

The mean-field free energy can be expressed as
\begin{equation}
	\mathcal{F}[\Delta_{\pm \mathbf{Q}_\alpha}] \equiv
	-\frac{1}{16\beta}\sum_{n,\bk}\mathrm{Tr}\,\mathrm{ln}\mathcal{G}^{-1}(i \omega_{n}, \bk)=\mathcal{F}^{(0)} -\frac{1}{16\beta}\sum_{n,\bk}\mathrm{Tr}\,\mathrm{ln}\left(
	1 + \mathcal{G}_{0}(i \omega_{n}, \bk)\Sigma(\bk)
	\right)
	=\mathcal{F}^{(0)} + \sum_{j=1}^{\infty}\mathcal{F}^{(2j)},
\end{equation}
where $\mathcal{F}^{(0)}$ is a constant that is independent of $\Delta_{\pm \mathbf{Q}_{\alpha}}$. $\mathcal{F}^{(2j)}$ is the free energy of order $|\Delta_{\pm \mathbf{Q}_{\alpha}}|^{2j}$ of the following form,
\begin{equation}
	\mathcal{F}^{(2j)} = \frac{1}{32j\beta}\sum_{n,\bk}\mathrm{Tr}
	\left[
	\left(
	\mathcal{G}_{0}(i \omega_n,\bk)\Sigma (\bk)
	\right)^{2j}
	\right]
	=\frac{1}{16j\beta}\sum_{n,\bk}\mathrm{Tr}\left[
	\left(
	-G_{0}(i \omega_n,\bk)\hat{\Delta}(\bk)G_{0}(-i\omega_n,-\bk)\hat{\Delta}^{\dagger}(\bk)
	\right)^{j}
	\right].
\end{equation}
Notice that the factor $\frac{1}{16}$ comes from the $4 \times 4$ folding of the system and the summation over $\bk$ is performed in the first BZ.

For $j=1$,
\begin{equation}
	\mathcal{F}^{(2)}
	=-\frac{1}{16\beta}\sum_{n,\bk}\mathrm{Tr}\left[
	G_{0}(i \omega_n,\bk)\hat{\Delta}(\bk)G_{0}(-i\omega_n,-\bk)\hat{\Delta}^{\dagger}(\bk)
	\right]=\frac{1}{16\beta} \sum_{n,\bk,i}\frac{e^{-\frac{\left|\xi_{\bk_i}\right|}{\Lambda}}}{i \omega_n -\xi_{\bk_{i}}} a(i \omega_n,\bk_{i}),
\end{equation}
where $\Lambda$ is the energy cutoff and
\begin{equation}
a(i \omega_n,\bk_{i})=\sum_{\alpha=1}^{3}
\left(
\frac{e^{-\frac{\left|\xi_{-\bk_i+\mathbf{Q}_\alpha}\right|}{\Lambda}}}{i \omega_n +\xi_{-\bk_i+\mathbf{Q}_\alpha}}
\left|\Delta_{\mathbf{Q}_{\alpha}} \right|^{2}
+
\frac{e^{-\frac{\left|\xi_{-\bk_i-\mathbf{Q}_\alpha}\right|}{\Lambda}}}{i \omega_n +\xi_{-\bk_i-\mathbf{Q}_\alpha}}
\left|\Delta_{-\mathbf{Q}_{\alpha}} \right|^{2}
\right).
\end{equation}
By performing the summation over $i \omega_n$ and $\bk$, we can obtain the following $\mathcal{F}^{(2)}$,
\begin{equation}
\mathcal{F}^{(2)}=g^{(2)}\sum_{\alpha=1}^{3}
\left(
\left|\Delta_{\mathbf{Q}_{\alpha}} \right|^{2}+\left|\Delta_{-\mathbf{Q}_{\alpha}} \right|^{2}
\right),
\end{equation}
where $g^{(2)}$ is the corresponding coefficient.

For $j=2$,
\begin{equation}\label{seq:F4-0}
	\mathcal{F}^{(4)}
	=\frac{1}{32\beta}\sum_{n,\bk}\mathrm{Tr}\left[
	\left(
	G_{0}(i \omega_n,\bk)\hat{\Delta}(\bk)G_{0}(-i\omega_n,-\bk)\hat{\Delta}^{\dagger}(\bk)
	\right)^2
	\right]=\frac{1}{32\beta} \sum_{n,i,j}{\sum_{\bk}}'\frac{1}{
\left(i \omega_n -\xi_{\bk_{i}}\right)
\left(i \omega_n -\xi_{\bk_{j}}\right)
} b(i \omega_n,\bk_{i},\bk_{j}),
\end{equation}
where we introduce ${\sum'_{\bk}}$ to represent the summation over $\bk$ with the energy cutoff as follows,
\begin{equation}
{\sum_{\bk}}'\frac{1}{
		\left(i \omega_n -\xi_{\bk_{i}}\right)
		\left(i \omega_n -\xi_{\bk_{j}}\right)
		\left(i \omega_n +\xi_{\bk_{p}}\right)
		\left(i \omega_n +\xi_{\bk_{q}}\right)
	}=
{\sum_{\bk}}\frac{e^{-\frac{
		\left| \xi_{\bk_{i}}\right| + \left| \xi_{\bk_{j}}\right|
	+\left| \xi_{\bk_{p}}\right|
+\left| \xi_{\bk_{q}}\right|}{\Lambda}}}{
	\left(i \omega_n -\xi_{\bk_{i}}\right)
	\left(i \omega_n -\xi_{\bk_{j}}\right)
	\left(i \omega_n +\xi_{\bk_{p}}\right)
	\left(i \omega_n +\xi_{\bk_{q}}\right)
}.
\end{equation}
Since the system is of $4 \times 4$ folded, we can keep $i=1$ as well as $\bk_{i}=\bk$ and calculate the summation over $j$, $i\omega_n$ and $\bk$. Then the summation over $i$ will give rise to $16$ copies. Thus, we can obtain the following $\mathcal{F}^{(4)}$ according to Eq.~\eqref{seq:F4-0},
\begin{equation}\label{seq:F4}
	\mathcal{F}^{(4)}
	=\frac{1}{2\beta} \sum_{n,j}{\sum_{\bk}}'\frac{1}{
		\left(i \omega_n -\xi_{\bk}\right)
		\left(i \omega_n -\xi_{\bk_{j}}\right)
	} b(i \omega_n,\bk,\bk_{j}),
\end{equation}
where $b(i\omega_n,\bk,\bk_{j})$ reads
\begin{subequations}\label{seq:b}
	\begin{align}
b(i\omega_n,\bk,\bk)&=
\left[\sum_{\alpha=1}^{3}
\left(
\frac{\left|\Delta_{\mathbf{Q}_{\alpha}} \right|^{2}}{i \omega_n +\xi_{-\bk+\mathbf{Q}_\alpha}}
+
\frac{\left|\Delta_{-\mathbf{Q}_{\alpha}} \right|^{2}}{i \omega_n +\xi_{-\bk-\mathbf{Q}_\alpha}}
\right)
\right]^{2},\\
\begin{split}
	b(i\omega_n,\bk,\bk+\mathbf{Q}_1)&=
	\left(
	\frac{\Delta_{\mathbf{Q}_2}\Delta^{*}_{-\mathbf{Q}_3}}{
		i \omega_n +\xi_{-\bk+{\mathbf{Q}_2}}
	}
	+
	\frac{\Delta_{\mathbf{Q}_3}\Delta^{*}_{-\mathbf{Q}_2}}{
		i \omega_n +\xi_{-\bk+{\mathbf{Q}_3}}
	}
	\right)
	\left(
\frac{\Delta_{-\mathbf{Q}_3} \Delta^{*}_{\mathbf{Q}_2} }{
		i \omega_n +\xi_{-\bk+{\mathbf{Q}_2}}
}
+
\frac{\Delta_{-\mathbf{Q}_2}\Delta^{*}_{\mathbf{Q}_3} }{
i \omega_n +\xi_{-\bk+{\mathbf{Q}_3}}
}
\right)\\
	&=\frac{
		\left| \Delta_{\mathbf{Q}_2}\right|^2
		\left| \Delta_{-\mathbf{Q}_3}\right|^2
	}{\left(
		i \omega_n+\xi_{-\bk + \mathbf{Q}_2}
		\right)^{2}}+
	\frac{
		\left| \Delta_{-\mathbf{Q}_2}\right|^2
	\left| \Delta_{\mathbf{Q}_3}\right|^2
}{\left(
		i \omega_n+\xi_{-\bk +\mathbf{Q}_3}
		\right)^{2}}+
	\frac{\left(\Delta_{\mathbf{Q}_2}\Delta_{-\mathbf{Q}_2}
		\right)
	\left(\Delta_{\mathbf{Q}_3}\Delta_{-\mathbf{Q}_3}
	\right)^{*}+c.c.}{
	\left(
	i \omega_n+\xi_{-\bk + \mathbf{Q}_2}
	\right)
	\left(
	i \omega_n+\xi_{-\bk +\mathbf{Q}_3}
	\right)
}
	,
\end{split}\\
\begin{split}
	b(i\omega_n,\bk,\bk-\mathbf{Q}_1)&=
	\left(
	\frac{\Delta_{-\mathbf{Q}_2}\Delta^{*}_{\mathbf{Q}_3}}{
		i \omega_n +\xi_{-\bk-{\mathbf{Q}_2}}
	}
	+
	\frac{\Delta_{-\mathbf{Q}_3}\Delta^{*}_{\mathbf{Q}_2}}{
		i \omega_n +\xi_{-\bk-{\mathbf{Q}_3}}
	}
	\right)
	\left(
	\frac{\Delta_{\mathbf{Q}_3} \Delta^{*}_{-\mathbf{Q}_2} }{
		i \omega_n +\xi_{-\bk-{\mathbf{Q}_2}}
	}
	+
	\frac{\Delta_{\mathbf{Q}_2}\Delta^{*}_{-\mathbf{Q}_3} }{
		i \omega_n +\xi_{-\bk-{\mathbf{Q}_3}}
	}
	\right)\\
	&=\frac{
		\left| \Delta_{-\mathbf{Q}_2}\right|^2
		\left| \Delta_{\mathbf{Q}_3}\right|^2
	}{\left(
		i \omega_n+\xi_{-\bk - \mathbf{Q}_2}
		\right)^{2}}+
	\frac{
		\left| \Delta_{\mathbf{Q}_2}\right|^2
		\left| \Delta_{-\mathbf{Q}_3}\right|^2
	}{\left(
		i \omega_n+\xi_{-\bk -\mathbf{Q}_3}
		\right)^{2}}+
	\frac{\left(\Delta_{\mathbf{Q}_2}\Delta_{-\mathbf{Q}_2}
		\right)
		\left(\Delta_{\mathbf{Q}_3}\Delta_{-\mathbf{Q}_3}
		\right)^{*}+c.c.}{
		\left(
		i \omega_n+\xi_{-\bk - \mathbf{Q}_2}
		\right)
		\left(
		i \omega_n+\xi_{-\bk -\mathbf{Q}_3}
		\right)
	}
	,
\end{split}\\
\begin{split}
	b(i\omega_n,\bk,\bk+\mathbf{Q}_2)&=
	\left(
	\frac{\Delta_{\mathbf{Q}_1}\Delta^{*}_{-\mathbf{Q}_3}}{
		i \omega_n +\xi_{-\bk+{\mathbf{Q}_1}}
	}
	+
	\frac{\Delta_{\mathbf{Q}_3}\Delta^{*}_{-\mathbf{Q}_1}}{
		i \omega_n +\xi_{-\bk+{\mathbf{Q}_3}}
	}
	\right)
		\left(
	\frac{\Delta_{-\mathbf{Q}_3}\Delta^{*}_{\mathbf{Q}_1}}{
		i \omega_n +\xi_{-\bk+{\mathbf{Q}_1}}
	}
	+
	\frac{\Delta_{-\mathbf{Q}_1}\Delta^{*}_{\mathbf{Q}_3}}{
		i \omega_n +\xi_{-\bk+{\mathbf{Q}_3}}
	}
	\right)\\
	&=\frac{
		\left| \Delta_{\mathbf{Q}_1}\right|^2
		\left| \Delta_{-\mathbf{Q}_3}\right|^2
	}{\left(
		i \omega_n+\xi_{-\bk + \mathbf{Q}_1}
		\right)^{2}}+
	\frac{
		\left| \Delta_{-\mathbf{Q}_1}\right|^2
		\left| \Delta_{\mathbf{Q}_3}\right|^2
	}{\left(
		i \omega_n+\xi_{-\bk +\mathbf{Q}_3}
		\right)^{2}}+
	\frac{\left(\Delta_{\mathbf{Q}_1}\Delta_{-\mathbf{Q}_1}
		\right)
		\left(\Delta_{\mathbf{Q}_3}\Delta_{-\mathbf{Q}_3}
		\right)^{*}+c.c.}{
		\left(
		i \omega_n+\xi_{-\bk +\mathbf{Q}_1}
		\right)
		\left(
		i \omega_n+\xi_{-\bk +\mathbf{Q}_3}
		\right)
	}
	,
\end{split}\\
\begin{split}
	b(i\omega_n,\bk,\bk-\mathbf{Q}_2)&=
	\left(
	\frac{\Delta_{-\mathbf{Q}_1}\Delta^{*}_{\mathbf{Q}_3}}{
		i \omega_n +\xi_{-\bk-{\mathbf{Q}_1}}
	}
	+
	\frac{\Delta_{-\mathbf{Q}_3}\Delta^{*}_{\mathbf{Q}_1}}{
		i \omega_n +\xi_{-\bk-{\mathbf{Q}_3}}
	}
	\right)
	\left(
\frac{\Delta_{\mathbf{Q}_3}\Delta^{*}_{-\mathbf{Q}_1}}{
	i \omega_n +\xi_{-\bk-{\mathbf{Q}_1}}
}
+
\frac{\Delta_{\mathbf{Q}_1}\Delta^{*}_{-\mathbf{Q}_3}}{
	i \omega_n +\xi_{-\bk-{\mathbf{Q}_3}}
}
\right)\\
	&=\frac{
		\left| \Delta_{-\mathbf{Q}_1}\right|^2
		\left| \Delta_{\mathbf{Q}_3}\right|^2
	}{\left(
		i \omega_n+\xi_{-\bk -\mathbf{Q}_1}
		\right)^{2}}+
	\frac{
		\left| \Delta_{\mathbf{Q}_1}\right|^2
		\left| \Delta_{-\mathbf{Q}_3}\right|^2
	}{\left(
		i \omega_n+\xi_{-\bk -\mathbf{Q}_3}
		\right)^{2}}+
	\frac{\left(\Delta_{\mathbf{Q}_1}\Delta_{-\mathbf{Q}_1}
		\right)
		\left(\Delta_{\mathbf{Q}_3}\Delta_{-\mathbf{Q}_3}
		\right)^{*}+c.c.}{
		\left(
		i \omega_n+\xi_{-\bk -\mathbf{Q}_1}
		\right)
		\left(
		i \omega_n+\xi_{-\bk -\mathbf{Q}_3}
		\right)
	}
	,
\end{split}\\
\begin{split}
	b(i\omega_n,\bk,\bk+\mathbf{Q}_3)&=
	\left(
	\frac{\Delta_{\mathbf{Q}_1}\Delta^{*}_{-\mathbf{Q}_2}}{
		i \omega_n +\xi_{-\bk+{\mathbf{Q}_1}}
	}
	+
	\frac{\Delta_{\mathbf{Q}_2}\Delta^{*}_{-\mathbf{Q}_1}}{
		i \omega_n +\xi_{-\bk+{\mathbf{Q}_2}}
	}
	\right)
	\left(
\frac{\Delta_{-\mathbf{Q}_2}\Delta^{*}_{\mathbf{Q}_1}}{
	i \omega_n +\xi_{-\bk+{\mathbf{Q}_1}}
}
+
\frac{\Delta_{-\mathbf{Q}_1}\Delta^{*}_{\mathbf{Q}_2}}{
	i \omega_n +\xi_{-\bk+{\mathbf{Q}_2}}
}
\right)\\
	&=\frac{
		\left| \Delta_{\mathbf{Q}_1}\right|^2
		\left| \Delta_{-\mathbf{Q}_2}\right|^2
	}{\left(
		i \omega_n+\xi_{-\bk +\mathbf{Q}_1}
		\right)^{2}}+
	\frac{
		\left| \Delta_{-\mathbf{Q}_1}\right|^2
		\left| \Delta_{\mathbf{Q}_2}\right|^2
	}{\left(
		i \omega_n+\xi_{-\bk +\mathbf{Q}_2}
		\right)^{2}}+
	\frac{\left(\Delta_{\mathbf{Q}_1}\Delta_{-\mathbf{Q}_1}
		\right)
		\left(\Delta_{\mathbf{Q}_2}\Delta_{-\mathbf{Q}_2}
		\right)^{*}+c.c.}{
		\left(
		i \omega_n+\xi_{-\bk +\mathbf{Q}_1}
		\right)
		\left(
		i \omega_n+\xi_{-\bk +\mathbf{Q}_2}
		\right)
	}
	,
\end{split}\\
\begin{split}
	b(i\omega_n,\bk,\bk-\mathbf{Q}_3)&=
	\left(
\frac{\Delta_{-\mathbf{Q}_1}\Delta^{*}_{\mathbf{Q}_2}}{
	i \omega_n +\xi_{-\bk-{\mathbf{Q}_1}}
}
+
\frac{\Delta_{-\mathbf{Q}_2}\Delta^{*}_{\mathbf{Q}_1}}{
	i \omega_n +\xi_{-\bk-{\mathbf{Q}_2}}
}
\right)
	\left(
\frac{\Delta_{\mathbf{Q}_2}\Delta^{*}_{-\mathbf{Q}_1}}{
	i \omega_n +\xi_{-\bk-{\mathbf{Q}_1}}
}
+
\frac{\Delta_{\mathbf{Q}_1}\Delta^{*}_{-\mathbf{Q}_2}}{
	i \omega_n +\xi_{-\bk-{\mathbf{Q}_2}}
}
\right)\\
	&=\frac{
		\left| \Delta_{-\mathbf{Q}_1}\right|^2
		\left| \Delta_{\mathbf{Q}_2}\right|^2
	}{\left(
		i \omega_n+\xi_{-\bk -\mathbf{Q}_1}
		\right)^{2}}+
	\frac{
		\left| \Delta_{\mathbf{Q}_1}\right|^2
		\left| \Delta_{-\mathbf{Q}_2}\right|^2
	}{\left(
		i \omega_n+\xi_{-\bk -\mathbf{Q}_2}
		\right)^{2}}+
	\frac{\left(\Delta_{\mathbf{Q}_1}\Delta_{-\mathbf{Q}_1}
		\right)
		\left(\Delta_{\mathbf{Q}_2}\Delta_{-\mathbf{Q}_2}
		\right)^{*}+c.c.}{
		\left(
		i \omega_n+\xi_{-\bk -\mathbf{Q}_1}
		\right)
		\left(
		i \omega_n+\xi_{-\bk -\mathbf{Q}_2}
		\right)
	}
	,
\end{split}\\
\begin{split}
	b(i\omega_n,\bk,\bk+2\mathbf{Q}_1)&=
	\left(
	\frac{\Delta_{\mathbf{Q}_1}\Delta^{*}_{-\mathbf{Q}_1}}{
		i \omega_n +\xi_{-\bk+{\mathbf{Q}_1}}
	}
	+
	\frac{\Delta_{-\mathbf{Q}_1}\Delta^{*}_{\mathbf{Q}_1}}{
		i \omega_n +\xi_{-\bk-{\mathbf{Q}_1}}
	}
	\right)
	\left(
\frac{\Delta_{-\mathbf{Q}_1}\Delta^{*}_{\mathbf{Q}_1}}{
	i \omega_n +\xi_{-\bk+{\mathbf{Q}_1}}
}
+
\frac{\Delta_{\mathbf{Q}_1}\Delta^{*}_{-\mathbf{Q}_1}}{
	i \omega_n +\xi_{-\bk-{\mathbf{Q}_1}}
}
\right)\\
	&=\frac{
		\left| \Delta_{\mathbf{Q}_1}\right|^2
		\left| \Delta_{-\mathbf{Q}_1}\right|^2
	}{\left(
		i \omega_n+\xi_{-\bk +\mathbf{Q}_1}
		\right)^{2}}+
	\frac{
		\left| \Delta_{\mathbf{Q}_1}\right|^2
		\left| \Delta_{-\mathbf{Q}_1}\right|^2
	}{\left(
		i \omega_n+\xi_{-\bk -\mathbf{Q}_1}
		\right)^{2}}+
	\frac{\left(\Delta^{2}_{\mathbf{Q}_1}
		\right)
		\left(\Delta^{2}_{-\mathbf{Q}_1}
		\right)^{*}+c.c.}{
		\left(
		i \omega_n+\xi_{-\bk +\mathbf{Q}_1}
		\right)
		\left(
		i \omega_n+\xi_{-\bk -\mathbf{Q}_1}
		\right)
	}
	,
\end{split}\\
\begin{split}
	b(i\omega_n,\bk,\bk+2\mathbf{Q}_2)&=
	\left(
	\frac{\Delta_{\mathbf{Q}_2}\Delta^{*}_{-\mathbf{Q}_2}}{
		i \omega_n +\xi_{-\bk+{\mathbf{Q}_2}}
	}
	+
	\frac{\Delta_{-\mathbf{Q}_2}\Delta^{*}_{\mathbf{Q}_2}}{
		i \omega_n +\xi_{-\bk-{\mathbf{Q}_2}}
	}
	\right)
	\left(
	\frac{\Delta_{-\mathbf{Q}_2}\Delta^{*}_{\mathbf{Q}_2}}{
		i \omega_n +\xi_{-\bk+{\mathbf{Q}_2}}
	}
	+
	\frac{\Delta_{\mathbf{Q}_2}\Delta^{*}_{-\mathbf{Q}_2}}{
		i \omega_n +\xi_{-\bk-{\mathbf{Q}_2}}
	}
	\right)\\
	&=\frac{
		\left| \Delta_{\mathbf{Q}_2}\right|^2
		\left| \Delta_{-\mathbf{Q}_2}\right|^2
	}{\left(
		i \omega_n+\xi_{-\bk +\mathbf{Q}_2}
		\right)^{2}}+
	\frac{
		\left| \Delta_{\mathbf{Q}_2}\right|^2
		\left| \Delta_{-\mathbf{Q}_2}\right|^2
	}{\left(
		i \omega_n+\xi_{-\bk -\mathbf{Q}_2}
		\right)^{2}}+
	\frac{\left(\Delta^{2}_{\mathbf{Q}_2}
	\right)
	\left(\Delta^{2}_{-\mathbf{Q}_2}
	\right)^{*}+c.c.}{
	\left(
	i \omega_n+\xi_{-\bk +\mathbf{Q}_2}
	\right)
	\left(
	i \omega_n+\xi_{-\bk -\mathbf{Q}_2}
	\right)
}
	,
\end{split}\\
\begin{split}
	b(i\omega_n,\bk,\bk+2\mathbf{Q}_3)&=
	\left(
	\frac{\Delta_{\mathbf{Q}_3}\Delta^{*}_{-\mathbf{Q}_3}}{
		i \omega_n +\xi_{-\bk+{\mathbf{Q}_3}}
	}
	+
	\frac{\Delta_{-\mathbf{Q}_3}\Delta^{*}_{\mathbf{Q}_3}}{
		i \omega_n +\xi_{-\bk-{\mathbf{Q}_3}}
	}
	\right)
	\left(
	\frac{\Delta_{-\mathbf{Q}_3}\Delta^{*}_{\mathbf{Q}_3}}{
		i \omega_n +\xi_{-\bk+{\mathbf{Q}_3}}
	}
	+
	\frac{\Delta_{\mathbf{Q}_3}\Delta^{*}_{-\mathbf{Q}_3}}{
		i \omega_n +\xi_{-\bk-{\mathbf{Q}_3}}
	}
	\right)\\
	&=\frac{
		\left| \Delta_{\mathbf{Q}_3}\right|^2
		\left| \Delta_{-\mathbf{Q}_3}\right|^2
	}{\left(
		i \omega_n+\xi_{-\bk +\mathbf{Q}_3}
		\right)^{2}}+
	\frac{
		\left| \Delta_{\mathbf{Q}_3}\right|^2
		\left| \Delta_{-\mathbf{Q}_3}\right|^2
	}{\left(
		i \omega_n+\xi_{-\bk -\mathbf{Q}_3}
		\right)^{2}}+
	\frac{\left(\Delta^{2}_{\mathbf{Q}_3}
		\right)
		\left(\Delta^{2}_{-\mathbf{Q}_3}
		\right)^{*}+c.c.}{
		\left(
		i \omega_n+\xi_{-\bk +\mathbf{Q}_3}
		\right)
		\left(
		i \omega_n+\xi_{-\bk -\mathbf{Q}_3}
		\right)
	}
	,
\end{split}\\
\begin{split}
	b(i\omega_n,\bk,\bk+\mathbf{Q}_1-\mathbf{Q}_2)&=
	\left(
	\frac{\Delta_{-\mathbf{Q}_1}\Delta^{*}_{-\mathbf{Q}_2}}{
		i \omega_n +\xi_{-\bk-{\mathbf{Q}_1}}
	}
	+
	\frac{\Delta_{\mathbf{Q}_2}\Delta^{*}_{\mathbf{Q}_1}}{
		i \omega_n +\xi_{-\bk+{\mathbf{Q}_2}}
	}
	\right)
		\left(
	\frac{\Delta_{-\mathbf{Q}_2}\Delta^{*}_{-\mathbf{Q}_1}}{
		i \omega_n +\xi_{-\bk-{\mathbf{Q}_1}}
	}
	+
	\frac{\Delta_{\mathbf{Q}_1}\Delta^{*}_{\mathbf{Q}_2}}{
		i \omega_n +\xi_{-\bk+{\mathbf{Q}_2}}
	}
	\right)\\
	&=\frac{
		\left| \Delta_{-\mathbf{Q}_1}\right|^2
		\left| \Delta_{-\mathbf{Q}_2}\right|^2
	}{\left(
		i \omega_n+\xi_{-\bk -\mathbf{Q}_1}
		\right)^{2}}+
	\frac{
		\left| \Delta_{\mathbf{Q}_1}\right|^2
		\left| \Delta_{\mathbf{Q}_2}\right|^2
	}{\left(
		i \omega_n+\xi_{-\bk +\mathbf{Q}_2}
		\right)^{2}}+
	\frac{\left(\Delta_{\mathbf{Q}_1}\Delta_{-\mathbf{Q}_1}
	\right)
	\left(\Delta_{\mathbf{Q}_2}\Delta_{-\mathbf{Q}_2}
	\right)^{*}+c.c.}{
	\left(
	i \omega_n+\xi_{-\bk -\mathbf{Q}_1}
	\right)
	\left(
	i \omega_n+\xi_{-\bk +\mathbf{Q}_2}
	\right)
}
	,
\end{split}\\
\begin{split}
	b(i\omega_n,\bk,\bk-\mathbf{Q}_1+\mathbf{Q}_2)&=
	\left(
\frac{\Delta_{\mathbf{Q}_1}\Delta^{*}_{\mathbf{Q}_2}}{
	i \omega_n +\xi_{-\bk+{\mathbf{Q}_1}}
}
+
\frac{\Delta_{-\mathbf{Q}_2}\Delta^{*}_{-\mathbf{Q}_1}}{
	i \omega_n +\xi_{-\bk-{\mathbf{Q}_2}}
}
\right)
\left(
\frac{\Delta_{\mathbf{Q}_2}\Delta^{*}_{\mathbf{Q}_1}}{
	i \omega_n +\xi_{-\bk+{\mathbf{Q}_1}}
}
+
\frac{\Delta_{-\mathbf{Q}_1}\Delta^{*}_{-\mathbf{Q}_2}}{
	i \omega_n +\xi_{-\bk-{\mathbf{Q}_2}}
}
\right)\\
	&=\frac{
		\left| \Delta_{\mathbf{Q}_1}\right|^2
		\left| \Delta_{\mathbf{Q}_2}\right|^2
	}{\left(
		i \omega_n+\xi_{-\bk+\mathbf{Q}_1}
		\right)^{2}}+
	\frac{
		\left| \Delta_{-\mathbf{Q}_1}\right|^2
		\left| \Delta_{-\mathbf{Q}_2}\right|^2
	}{\left(
		i \omega_n+\xi_{-\bk +\mathbf{Q}_2}
		\right)^{2}}+
	\frac{\left(\Delta_{\mathbf{Q}_1}\Delta_{-\mathbf{Q}_1}
		\right)
		\left(\Delta_{\mathbf{Q}_2}\Delta_{-\mathbf{Q}_2}
		\right)^{*}+c.c.}{
		\left(
		i \omega_n+\xi_{-\bk +\mathbf{Q}_1}
		\right)
		\left(
		i \omega_n+\xi_{-\bk -\mathbf{Q}_2}
		\right)
	}
	,
\end{split}\\
\begin{split}
	b(i\omega_n,\bk,\bk+\mathbf{Q}_1-\mathbf{Q}_3)&=
	\left(
	\frac{\Delta_{-\mathbf{Q}_1}\Delta^{*}_{-\mathbf{Q}_3}} {
		i \omega_n +\xi_{-\bk-{\mathbf{Q}_1}}
	}
	+
	\frac{\Delta_{\mathbf{Q}_3}\Delta^{*}_{\mathbf{Q}_1}}{
		i \omega_n +\xi_{-\bk+{\mathbf{Q}_3}}
	}
	\right)
		\left(
	\frac{\Delta_{-\mathbf{Q}_3}\Delta^{*}_{-\mathbf{Q}_1}} {
		i \omega_n +\xi_{-\bk-{\mathbf{Q}_1}}
	}
	+
	\frac{\Delta_{\mathbf{Q}_1}\Delta^{*}_{\mathbf{Q}_3}}{
		i \omega_n +\xi_{-\bk+{\mathbf{Q}_3}}
	}
	\right)\\
	&=\frac{
		\left| \Delta_{-\mathbf{Q}_1}\right|^2
		\left| \Delta_{-\mathbf{Q}_3}\right|^2
	}{\left(
		i \omega_n+\xi_{-\bk-\mathbf{Q}_1}
		\right)^{2}}+
	\frac{
		\left| \Delta_{\mathbf{Q}_1}\right|^2
		\left| \Delta_{\mathbf{Q}_3}\right|^2
	}{\left(
		i \omega_n+\xi_{-\bk +\mathbf{Q}_3}
		\right)^{2}}+
	\frac{\left(\Delta_{\mathbf{Q}_1}\Delta_{-\mathbf{Q}_1}
		\right)
		\left(\Delta_{\mathbf{Q}_3}\Delta_{-\mathbf{Q}_3}
		\right)^{*}+c.c.}{
		\left(
		i \omega_n+\xi_{-\bk -\mathbf{Q}_1}
		\right)
		\left(
		i \omega_n+\xi_{-\bk +\mathbf{Q}_3}
		\right)
	}
	,
\end{split}\\
\begin{split}
	b(i\omega_n,\bk,\bk-\mathbf{Q}_1+\mathbf{Q}_3)&=
	\left(
	\frac{\Delta_{\mathbf{Q}_1}\Delta^{*}_{\mathbf{Q}_3}} {
		i \omega_n +\xi_{-\bk+{\mathbf{Q}_1}}
	}
	+
	\frac{\Delta_{-\mathbf{Q}_3}\Delta^{*}_{-\mathbf{Q}_1}}{
		i \omega_n +\xi_{-\bk-{\mathbf{Q}_3}}
	}
	\right)
	\left(
	\frac{\Delta_{\mathbf{Q}_3}\Delta^{*}_{\mathbf{Q}_1}} {
		i \omega_n +\xi_{-\bk+{\mathbf{Q}_1}}
	}
	+
	\frac{\Delta_{-\mathbf{Q}_1}\Delta^{*}_{-\mathbf{Q}_3}}{
		i \omega_n +\xi_{-\bk-{\mathbf{Q}_3}}
	}
	\right)\\
	&=\frac{
		\left| \Delta_{\mathbf{Q}_1}\right|^2
		\left| \Delta_{\mathbf{Q}_3}\right|^2
	}{\left(
		i \omega_n+\xi_{-\bk+\mathbf{Q}_1}
		\right)^{2}}+
	\frac{
		\left| \Delta_{-\mathbf{Q}_1}\right|^2
		\left| \Delta_{-\mathbf{Q}_3}\right|^2
	}{\left(
		i \omega_n+\xi_{-\bk -\mathbf{Q}_3}
		\right)^{2}}+
	\frac{\left(\Delta_{\mathbf{Q}_1}\Delta_{-\mathbf{Q}_1}
		\right)
		\left(\Delta_{\mathbf{Q}_3}\Delta_{-\mathbf{Q}_3}
		\right)^{*}+c.c.}{
		\left(
		i \omega_n+\xi_{-\bk +\mathbf{Q}_1}
		\right)
		\left(
		i \omega_n+\xi_{-\bk -\mathbf{Q}_3}
		\right)
	}
	,
\end{split}\\
\begin{split}
	b(i\omega_n,\bk,\bk+\mathbf{Q}_2-\mathbf{Q}_3)&=
	\left(
	\frac{\Delta_{-\mathbf{Q}_2}\Delta^{*}_{-\mathbf{Q}_3}} {
		i \omega_n +\xi_{-\bk-{\mathbf{Q}_2}}
	}
	+
	\frac{\Delta_{\mathbf{Q}_3}\Delta^{*}_{\mathbf{Q}_2}}{
		i \omega_n +\xi_{-\bk+{\mathbf{Q}_3}}
	}
	\right)
	\left(
\frac{\Delta_{-\mathbf{Q}_3}\Delta^{*}_{-\mathbf{Q}_2}} {
	i \omega_n +\xi_{-\bk-{\mathbf{Q}_2}}
}
+
\frac{\Delta_{\mathbf{Q}_2}\Delta^{*}_{\mathbf{Q}_3}}{
	i \omega_n +\xi_{-\bk+{\mathbf{Q}_3}}
}
\right)\\
	&=\frac{
		\left| \Delta_{-\mathbf{Q}_2}\right|^2
		\left| \Delta_{-\mathbf{Q}_3}\right|^2
	}{\left(
		i \omega_n+\xi_{-\bk-\mathbf{Q}_2}
		\right)^{2}}+
	\frac{
		\left| \Delta_{\mathbf{Q}_2}\right|^2
		\left| \Delta_{\mathbf{Q}_3}\right|^2
	}{\left(
		i \omega_n+\xi_{-\bk +\mathbf{Q}_3}
		\right)^{2}}+
	\frac{\left(\Delta_{\mathbf{Q}_2}\Delta_{-\mathbf{Q}_2}
		\right)
		\left(\Delta_{\mathbf{Q}_3}\Delta_{-\mathbf{Q}_3}
		\right)^{*}+c.c.}{
		\left(
		i \omega_n+\xi_{-\bk -\mathbf{Q}_2}
		\right)
		\left(
		i \omega_n+\xi_{-\bk +\mathbf{Q}_3}
		\right)
	}
	,
\end{split}\\
\begin{split}
	b(i\omega_n,\bk,\bk-\mathbf{Q}_2+\mathbf{Q}_3)&=
	\left(
	\frac{\Delta_{\mathbf{Q}_2}\Delta^{*}_{\mathbf{Q}_3}} {
		i \omega_n +\xi_{-\bk+{\mathbf{Q}_2}}
	}
	+
	\frac{\Delta_{-\mathbf{Q}_3}\Delta^{*}_{-\mathbf{Q}_2}}{
		i \omega_n +\xi_{-\bk-{\mathbf{Q}_3}}
	}
	\right)
	\left(
	\frac{\Delta_{\mathbf{Q}_3}\Delta^{*}_{\mathbf{Q}_2}} {
		i \omega_n +\xi_{-\bk+{\mathbf{Q}_2}}
	}
	+
	\frac{\Delta_{-\mathbf{Q}_2}\Delta^{*}_{-\mathbf{Q}_3}}{
		i \omega_n +\xi_{-\bk-{\mathbf{Q}_3}}
	}
	\right)\\
	&=\frac{
		\left| \Delta_{\mathbf{Q}_2}\right|^2
		\left| \Delta_{\mathbf{Q}_3}\right|^2
	}{\left(
		i \omega_n+\xi_{-\bk+\mathbf{Q}_2}
		\right)^{2}}+
	\frac{
		\left| \Delta_{-\mathbf{Q}_2}\right|^2
		\left| \Delta_{-\mathbf{Q}_3}\right|^2
	}{\left(
		i \omega_n+\xi_{-\bk -\mathbf{Q}_3}
		\right)^{2}}+
	\frac{\left(\Delta_{\mathbf{Q}_2}\Delta_{-\mathbf{Q}_2}
		\right)
		\left(\Delta_{\mathbf{Q}_3}\Delta_{-\mathbf{Q}_3}
		\right)^{*}+c.c.}{
		\left(
		i \omega_n+\xi_{-\bk +\mathbf{Q}_2}
		\right)
		\left(
		i \omega_n+\xi_{-\bk -\mathbf{Q}_3}
		\right)
	}.
\end{split}
\end{align}
\end{subequations}
By performing the summation over $i \omega_n$ and $\bk$, $\mathcal{F}^{(4)}$ is of the following form,
\begin{equation}\label{seq:F4f}
	\begin{aligned}
	\mathcal{F}^{(4)}=&g^{(4)}_{1}
	\left(
	\left| \Delta_{\mathbf{Q}_1}\right|^{4}+\left|\Delta_{-\mathbf{Q}_1}\right|^{4}
	+\left| \Delta_{\mathbf{Q}_2}\right|^{4}+\left|\Delta_{-\mathbf{Q}_2}\right|^{4}
	+\left| \Delta_{\mathbf{Q}_3}\right|^{4}+\left|\Delta_{-\mathbf{Q}_3}\right|^{4}
	\right)\\
	+&g^{(4)}_{2}
	\left(
	\left| \Delta_{\mathbf{Q}_1}\right|^{2} \left|\Delta_{-\mathbf{Q}_1}\right|^{2}
	+\left| \Delta_{\mathbf{Q}_2}\right|^{2} \left|\Delta_{-\mathbf{Q}_2}\right|^{2}
	+\left| \Delta_{\mathbf{Q}_3}\right|^{2} \left|\Delta_{-\mathbf{Q}_3}\right|^{2}
	\right)\\
	+&g^{(4)}_{3}
   	\left(
    \left| \Delta_{\mathbf{Q}_1}\right|^{2} \left|\Delta_{\mathbf{Q}_2}\right|^{2}
    +\left| \Delta_{\mathbf{Q}_2}\right|^{2} \left|\Delta_{\mathbf{Q}_3}\right|^{2}
    +\left| \Delta_{\mathbf{Q}_3}\right|^{2} \left|\Delta_{\mathbf{Q}_1}\right|^{2}
    +\left| \Delta_{-\mathbf{Q}_1}\right|^{2} \left|\Delta_{-\mathbf{Q}_2}\right|^{2}
    +\left| \Delta_{-\mathbf{Q}_2}\right|^{2} \left|\Delta_{-\mathbf{Q}_3}\right|^{2}
    +\left| \Delta_{-\mathbf{Q}_3}\right|^{2} \left|\Delta_{-\mathbf{Q}_1}\right|^{2}
    \right)\\
    +&g^{(4)}_{4}
    \left(
    \left| \Delta_{\mathbf{Q}_1}\right|^{2} \left|\Delta_{-\mathbf{Q}_2}\right|^{2}
    +\left| \Delta_{\mathbf{Q}_2}\right|^{2} \left|\Delta_{-\mathbf{Q}_3}\right|^{2}
    +\left| \Delta_{\mathbf{Q}_3}\right|^{2} \left|\Delta_{-\mathbf{Q}_1}\right|^{2}
    +\left| \Delta_{-\mathbf{Q}_1}\right|^{2} \left|\Delta_{\mathbf{Q}_2}\right|^{2}
    +\left| \Delta_{-\mathbf{Q}_2}\right|^{2} \left|\Delta_{\mathbf{Q}_3}\right|^{2}
    +\left| \Delta_{-\mathbf{Q}_3}\right|^{2} \left|\Delta_{\mathbf{Q}_1}\right|^{2}
    \right)\\
    +&g^{(4)}_{\phi}
    \left[
\left(\Delta^{2}_{\mathbf{Q}_1}\right) \left(\Delta^{2}_{-\mathbf{Q}_1}
\right)^{*}
+\left(\Delta^{2}_{\mathbf{Q}_2}\right) \left(\Delta^{2}_{-\mathbf{Q}_2}
\right)^{*}
+\left(\Delta^{2}_{\mathbf{Q}_3}\right) \left(\Delta^{2}_{-\mathbf{Q}_3}
\right)^{*}+c.c.
    \right]\\
    +&g^{(4)}_{\theta}
    \left[
    \left(\Delta_{\mathbf{Q}_1}\Delta_{-\mathbf{Q}_1}
    \right)
    \left(\Delta_{\mathbf{Q}_2}\Delta_{-\mathbf{Q}_2}
    \right)^{*}
    +\left(\Delta_{\mathbf{Q}_2}\Delta_{-\mathbf{Q}_2}
    \right)
    \left(\Delta_{\mathbf{Q}_3}\Delta_{-\mathbf{Q}_3}
    \right)^{*}
    +\left(\Delta_{\mathbf{Q}_3}\Delta_{-\mathbf{Q}_3}
    \right)
    \left(\Delta_{\mathbf{Q}_1}\Delta_{-\mathbf{Q}_1}
    \right)^{*}+c.c.
    \right],
    \end{aligned}
\end{equation}
where $g^{(4)}_1,\,g^{(4)}_2,\,g^{(4)}_3,\,g^{(4)}_4,\,g^{(4)}_{\phi}$ and $g^{(4)}_{\theta}$ are the corresponding $\Delta_{\pm \mathbf{Q}_\alpha}$-independent coefficients. For $\Delta_{\pm \mathbf{Q}_\alpha}=\Delta e^{i \theta_\alpha} e^{\pm i \frac{\phi_{\alpha}}{2}}$, we have
\begin{equation}\label{seq:F4ff}
		\mathcal{F}^{(4)}=6\left(g^{(4)}_1+\frac{g^{(4)}_2}{2}+g^{(4)}_3+g^{(4)}_4\right)\Delta^{4}
		+2g^{(4)}_{\phi}\Delta^{4}\sum_{\alpha=1}^{3}\cos \left(2\phi_{\alpha}\right) + 2g^{(4)}_{\theta}\Delta^{4}\left[
		\cos \left(2\theta_2-2\theta_1\right)+\cos \left(2\theta_3-2\theta_2\right)+\cos \left(2\theta_1-2\theta_3\right)
		\right].
\end{equation}

We are interested in the signs of $g^{(4)}_{\phi}$ and $g^{(4)}_{\theta}$ since they determine the values of $\phi_{\alpha}$ and $\theta_\alpha$ that minimize the free energy respectively.  According to Eq.~\eqref{seq:F4} and Eqs.~\eqref{seq:b}, $g^{(4)}_{\phi}$ and $g^{(4)}_{\theta}$ can be expressed as 
\begin{subequations}
	\begin{align}
		g^{(4)}_{\phi}&=\frac{1}{2\beta}\sum_{n}{\sum_{\bk}}'\frac{1}{
			\left( i \omega_n-\xi_{\bk} \right)	\left( i \omega_n-\xi_{\bk+2\mathbf{Q}_1} \right)	
			\left( i \omega_n+\xi_{-\bk+\mathbf{Q}_1} \right)	\left( i \omega_n+\xi_{-\bk-\mathbf{Q}_1} \right)	
		},\\
		g^{(4)}_{\theta}&=\frac{1}{\beta}\sum_{n}{\sum_{\bk}}'\frac{1}{
		\left( i \omega_n-\xi_{\bk} \right)	\left( i \omega_n-\xi_{\bk-\mathbf{Q}_1} \right)	
		\left( i \omega_n+\xi_{-\bk-\mathbf{Q}_2} \right)	\left( i \omega_n+\xi_{-\bk-\mathbf{Q}_3} \right)	
	}
+
\frac{1}{
	\left( i \omega_n-\xi_{\bk} \right)	\left( i \omega_n-\xi_{\bk+\mathbf{Q}_2 -\mathbf{Q}_3}\right)	
	\left( i \omega_n+\xi_{-\bk-\mathbf{Q}_2} \right)	\left( i \omega_n+\xi_{-\bk+\mathbf{Q}_3} \right)	
},
	\end{align}
\end{subequations}
where we have used $\xi_{\bk}=\xi_{-\bk}$ to combine terms in $g^{(4)}_{\theta}$. Now we can use the following formula to achieve the summation over $i\omega_n$,
\begin{equation}
	\begin{aligned}
		 &\frac{1}{\beta}\sum_{n}\frac{1}{
			\left(i \omega_n-\xi_{\bk_i}\right)
			\left(i \omega_n-\xi_{\bk_j}\right)
			\left(i \omega_n+\xi_{\bk_p}\right)
			\left(i \omega_n+\xi_{\bk_q}\right)}\\
		=&\frac{n_F(\xi_{\bk_i})}{\left(\xi_{\bk_i}-\xi_{\bk_j}\right)
			\left(\xi_{\bk_i}+\xi_{\bk_p}\right)
			\left(\xi_{\bk_i}+\xi_{\bk_q}\right)}
		+\frac{n_F(\xi_{\bk_j})}{\left(\xi_{\bk_j}-\xi_{\bk_i}\right)
			\left(\xi_{\bk_j}+\xi_{\bk_p}\right)
			\left(\xi_{\bk_j}+\xi_{\bk_q}\right)}\\
			+&\frac{n_F(-\xi_{\bk_p})}{
			\left(\xi_{\bk_i}+\xi_{\bk_p}\right)
			\left(\xi_{\bk_j}+\xi_{\bk_p}\right)\left(\xi_{\bk_q}-\xi_{\bk_p}\right)}
		+\frac{n_F(-\xi_{\bk_q})}{
			\left(\xi_{\bk_i}+\xi_{\bk_q}\right)
			\left(\xi_{\bk_j}+\xi_{\bk_q}\right)\left(\xi_{\bk_p}-\xi_{\bk_q}\right)},
	\end{aligned}
\end{equation}
where $n_F$ is the Fermi-Dirac distribution function. 

By performing the summation over $i \omega_n$, $g^{(4)}_{\phi}$ is of the following form,
\begin{equation}
	\begin{aligned}
		g^{(4)}_{\phi}=\frac{1}{2}{\sum_{\bk}}'&
		\frac{n_F(\xi_{\bk})}{\left(\xi_{\bk}-\xi_{\bk+2\mathbf{Q}_1}\right)
			\left(\xi_{\bk}+\xi_{-\bk+\mathbf{Q}_1}\right)
			\left(\xi_{\bk}+\xi_{-\bk-\mathbf{Q}_1}\right)}+
		\frac{n_F(\xi_{\bk+2\mathbf{Q}_1})}{\left(\xi_{\bk+2\mathbf{Q}_1}-\xi_{\bk}\right)
		\left(\xi_{\bk+2\mathbf{Q}_1}+\xi_{-\bk+\mathbf{Q}_1}\right)
		\left(\xi_{\bk+2\mathbf{Q}_1}+\xi_{-\bk-\mathbf{Q}_1}\right)}\\
		+&\frac{n_F(-\xi_{-\bk+\mathbf{Q}_1})}{
			\left(\xi_{\bk}+\xi_{-\bk+\mathbf{Q}_1}\right)
			\left(\xi_{\bk+2\mathbf{Q}_1}+\xi_{-\bk+\mathbf{Q}_1}\right)\left(\xi_{-\bk-\mathbf{Q}_1}-\xi_{-\bk+\mathbf{Q}_1}\right)}
		+\frac{n_F(-\xi_{-\bk-\mathbf{Q}_1})}{
			\left(\xi_{\bk}+\xi_{-\bk-\mathbf{Q}_1}\right)
			\left(\xi_{\bk+2\mathbf{Q}_1}+\xi_{-\bk-\mathbf{Q}_1}\right)\left(\xi_{-\bk+\mathbf{Q}_1}-\xi_{-\bk-\mathbf{Q}_1}\right)}\\
		={\sum_{\bk}}'&\frac{2n_F(\xi_{\bk})-1}{\left(\xi_{\bk}-\xi_{\bk+2\mathbf{Q}_1}\right)
			\left(\xi_{\bk}+\xi_{-\bk+\mathbf{Q}_1}\right)
			\left(\xi_{\bk}+\xi_{-\bk-\mathbf{Q}_1}\right)},
	\end{aligned}
\end{equation}
where we have used $n_F(x)+n_F(-x)=1$. Due to the energy cutoff $\Lambda$, only the summation near the FS segment defined by $k_x=\pm\pi$ will give rise to a sizable contribution to the expression above. Notice that near $k_x=\pm\pi$, the following relation that represents the nesting feature of the hexagonal FS holds,
$$
\xi_{\bk+2\mathbf{Q}_1} \simeq -\xi_{\bk},\, 
\xi_{-\bk-\mathbf{Q}_1} \simeq -\xi_{-\bk+\mathbf{Q}_1},
$$ 
In the low-energy limit ($|\beta{}\xi_{\bk}| \ll1$), the function $2n_{F}(\xi_{\bk})-1$  can be expanded as follows
$$
2n_{F}(\xi_{\bk})-1 \simeq -\frac{\beta\xi_{\bk}}{2} + \frac{\beta^3\xi^{3}_{\bk}}{24}.
$$
Then we have 
\begin{equation*}
	g^{(4)}_{\phi} \simeq {\sum_{\bk}}'\frac{-\frac{\beta\xi_{\bk}}{2} + \frac{\beta^3\xi^{3}_{\bk}}{24}}{2\xi_\bk
		\left(\xi_{\bk}^{2}-\xi_{-\bk+\mathbf{Q}_1}^{2}\right)}={\sum_{\bk}}'
	-\frac{\beta}{4\left(\xi_{\bk}^{2}-\xi_{-\bk+\mathbf{Q}_1}^{2}\right)}
	+\frac{\beta^3\xi^2_{\bk}}{48\left(\xi_{\bk}^{2}-\xi_{-\bk+\mathbf{Q}_1}^{2}\right)}.
\end{equation*}
By taking the substitution $\bk \rightarrow -\bk +\mathbf{Q}_1$ and adding it to the original expression, we obtain the following expression,
\begin{equation}
	g^{(4)}_{\phi} \simeq {\sum_{\bk}}'
	\frac{\beta^3\left(\xi_{\bk}^{2}-\xi_{-\bk+\mathbf{Q}_1}^{2}\right)}{96\left(\xi_{\bk}^{2}-\xi_{-\bk+\mathbf{Q}_1}^{2}\right)}={\sum_{\bk}}'\frac{\beta^3}{96} >0.
\end{equation}
We draw the conclusion that $g^{(4)}_{\phi}$ is positive.

We now focus on the derivation of $g^{(4)}_{\theta}$. Similarly, we perform the summation over $i \omega_n$ at first, 
\begin{subequations}
\begin{equation}
	\begin{aligned}
	&\frac{1}{\beta}\sum_{n}{\sum_{\bk}}'
	\frac{1}{
		\left( i \omega_n-\xi_{\bk} \right)	\left( i \omega_n-\xi_{\bk-\mathbf{Q}_1} \right)	
		\left( i \omega_n+\xi_{-\bk-\mathbf{Q}_2} \right)	\left( i \omega_n+\xi_{-\bk-\mathbf{Q}_3} \right)}	\\
		=&{\sum_{\bk}}'
		\frac{n_F(\xi_{\bk})}{\left(\xi_{\bk}-\xi_{\bk-\mathbf{Q}_1}\right)
			\left(\xi_{\bk}+\xi_{-\bk-\mathbf{Q}_2}\right)
			\left(\xi_{\bk}+\xi_{-\bk-\mathbf{Q}_3}\right)}+
		\frac{n_F(\xi_{\bk-\mathbf{Q}_1})}{\left(\xi_{\bk-\mathbf{Q}_1}-\xi_{\bk}\right)
			\left(\xi_{\bk-\mathbf{Q}_1}+\xi_{-\bk-\mathbf{Q}_2}\right)
			\left(\xi_{\bk-\mathbf{Q}_1}+\xi_{-\bk-\mathbf{Q}_3}\right)}\\
		+&\frac{n_F(-\xi_{-\bk-\mathbf{Q}_2})}{
			\left(\xi_{\bk}+\xi_{-\bk-\mathbf{Q}_2}\right)
			\left(\xi_{\bk-\mathbf{Q}_1}+\xi_{-\bk-\mathbf{Q}_2}\right)\left(\xi_{-\bk-\mathbf{Q}_3}-\xi_{-\bk-\mathbf{Q}_2}\right)}
		+\frac{n_F(-\xi_{-\bk-\mathbf{Q}_3})}{
			\left(\xi_{\bk}+\xi_{-\bk-\mathbf{Q}_3}\right)
			\left(\xi_{\bk-\mathbf{Q}_1}+\xi_{-\bk-\mathbf{Q}_3}\right)\left(\xi_{-\bk-\mathbf{Q}_2}-\xi_{-\bk-\mathbf{Q}_3}\right)}\\
		=&2{\sum_{\bk}}'
		\frac{n_F(\xi_{\bk})}{\left(\xi_{\bk}-\xi_{\bk-\mathbf{Q}_1}\right)
			\left(\xi_{\bk}+\xi_{-\bk-\mathbf{Q}_2}\right)
			\left(\xi_{\bk}+\xi_{-\bk-\mathbf{Q}_3}\right)}
		-	\frac{n_F(-\xi_{\bk})}{\left(\xi_{\bk}-\xi_{\bk+\mathbf{Q}_3-\mathbf{Q}_2}\right)
			\left(\xi_{\bk}+\xi_{-\bk+\mathbf{Q}_2}\right)
			\left(\xi_{\bk}+\xi_{-\bk-\mathbf{Q}_3}\right)}
\end{aligned},
\end{equation}
\begin{equation}
	\begin{aligned}
		&\frac{1}{\beta}\sum_{n}{\sum_{\bk}}'
		\frac{1}{
			\left( i \omega_n-\xi_{\bk} \right)	\left( i \omega_n-\xi_{\bk+\mathbf{Q}_2-\mathbf{Q}_3} \right)	
			\left( i \omega_n+\xi_{-\bk-\mathbf{Q}_2} \right)	\left( i \omega_n+\xi_{-\bk+\mathbf{Q}_3} \right)}	\\
		=&{\sum_{\bk}}'
		\frac{n_F(\xi_{\bk})}{\left(\xi_{\bk}-\xi_{\bk+\mathbf{Q}_2-\mathbf{Q}_3}\right)
			\left(\xi_{\bk}+\xi_{-\bk-\mathbf{Q}_2}\right)
			\left(\xi_{\bk}+\xi_{-\bk+\mathbf{Q}_3}\right)}+
		\frac{n_F(\xi_{\bk+\mathbf{Q}_2-\mathbf{Q}_3})}{\left(\xi_{\bk+\mathbf{Q}_2-\mathbf{Q}_3}-\xi_{\bk}\right)
			\left(\xi_{\bk+\mathbf{Q}_2-\mathbf{Q}_3}+\xi_{-\bk-\mathbf{Q}_2}\right)
			\left(\xi_{\bk+\mathbf{Q}_2-\mathbf{Q}_3}+\xi_{-\bk+\mathbf{Q}_3}\right)}\\
		+&\frac{n_F(-\xi_{-\bk-\mathbf{Q}_2})}{
			\left(\xi_{\bk}+\xi_{-\bk-\mathbf{Q}_2}\right)
			\left(\xi_{\bk+\mathbf{Q}_2-\mathbf{Q}_3}+\xi_{-\bk-\mathbf{Q}_2}\right)\left(\xi_{-\bk+\mathbf{Q}_3}-\xi_{-\bk-\mathbf{Q}_2}\right)}
		+\frac{n_F(-\xi_{-\bk+\mathbf{Q}_3})}{
			\left(\xi_{\bk}+\xi_{-\bk+\mathbf{Q}_3}\right)
			\left(\xi_{\bk+\mathbf{Q}_2-\mathbf{Q}_3}+\xi_{-\bk+\mathbf{Q}_3}\right)\left(\xi_{-\bk-\mathbf{Q}_2}-\xi_{-\bk+\mathbf{Q}_3}\right)}\\
		=&2{\sum_{\bk}}'
			\frac{n_F(\xi_{\bk})}{\left(\xi_{\bk}-\xi_{\bk+\mathbf{Q}_2-\mathbf{Q}_3}\right)
			\left(\xi_{\bk}+\xi_{-\bk-\mathbf{Q}_2}\right)
			\left(\xi_{\bk}+\xi_{-\bk+\mathbf{Q}_3}\right)}
		-	\frac{n_F(-\xi_{\bk})}{\left(\xi_{\bk}-\xi_{\bk+\mathbf{Q}_1}\right)
			\left(\xi_{\bk}+\xi_{-\bk+\mathbf{Q}_2}\right)
			\left(\xi_{\bk}+\xi_{-\bk+\mathbf{Q}_3}\right)}
	\end{aligned}.
\end{equation}
\end{subequations}
Then $g^{(4)}_{\theta}$ is of the following form,
\begin{subequations}
\begin{align}
  g^{(4)}_{\theta}&=g^{(4)}_{\theta,1} + g^{(4)}_{\theta,2},\\
  g^{(4)}_{\theta,1}&= 2{\sum_{\bk}}'
  \frac{2n_F(\xi_{\bk})-1}{\left(\xi_{\bk}-\xi_{\bk+\mathbf{Q}_2-\mathbf{Q}_3}\right)
  	\left(\xi_{\bk}+\xi_{-\bk-\mathbf{Q}_2}\right)
  	\left(\xi_{\bk}+\xi_{-\bk+\mathbf{Q}_3}\right)}             ,\\
   g^{(4)}_{\theta,2}&=2{\sum_{\bk}}'
   \frac{2n_F(\xi_{\bk})-1}{\left(\xi_{\bk}-\xi_{\bk+\mathbf{Q}_1}\right)
   	\left(\xi_{\bk}+\xi_{-\bk+\mathbf{Q}_2}\right)
   	\left(\xi_{\bk}+\xi_{-\bk+\mathbf{Q}_3}\right)}.
\end{align}
\end{subequations}
For $g^{(4)}_{\theta,1}$, only $\bk$ near $\left(0,2\pi/\sqrt{3}\right)$ or  $\left(-\pi, 0\right)$ gives a sizable contribution to the summation above due to the energy cutoff. For $\bk=\left(0,2\pi/\sqrt{3}\right)+\bq$, we have
$$
\xi_{\bk} \simeq \frac{1}{2}(q^2_x-3q^2_y),\,
\xi_{\bk+\mathbf{Q}_2-\mathbf{Q}_3} \simeq 2q_x,\,
\xi_{-\bk-\mathbf{Q}_2} \simeq -q_x+\sqrt{3}q_y,\,
\xi_{-\bk+\mathbf{Q}_3} \simeq -q_x-\sqrt{3}q_y,
$$
where $\bq$ is a small vector. Similarly, for $\bk=\left(-\pi, 0\right)+\bq$, we have
$$
\xi_{\bk} \simeq -2q_x,\,
\xi_{\bk+\mathbf{Q}_2-\mathbf{Q}_3} \simeq \frac{1}{2}(q^2_x-3q^2_y),\,
\xi_{-\bk-\mathbf{Q}_2} \simeq q_x+\sqrt{3}q_y,\,
\xi_{-\bk+\mathbf{Q}_3} \simeq q_x-\sqrt{3}q_y.
$$
Thus we can obtain the following approximation of $g^{(4)}_{\theta,1}$,
\begin{equation}
	\begin{aligned}
	g^{(4)}_{\theta,1} &\simeq
	\sum_{\bq} \frac{2n_F\left(\frac{1}{2}(q^2_x-3q^2_y)\right)-1}{-q_x
		\left(-q_x+\sqrt{3}q_y
		\right)
		\left(-q_x-\sqrt{3}q_y
		\right)
	}
	+
\frac{-\beta(-2q_x)+\frac{\beta^3(-2q_x)^3}{12}}{-2q_x
	\left(-q_x+\sqrt{3}q_y
	\right)
	\left(-q_x-\sqrt{3}q_y
	\right)}\\
&=\sum_{\bq}-\frac{\beta}{q^2_x-3q^2_y}+\frac{\beta^3q^2_x}{3\left(q^2_x-3q^2_y\right)},
	\end{aligned}
\end{equation}
where the first term in the first line vanishes since it changes sign under $\bq \rightarrow -\bq$.

For $g^{(4)}_{\theta,2}$, only $\bk$ near $\left(\pm\pi/2,\sqrt{3}\pi/2\right)$ gives a sizable contribution to the summation above. For $\bk=\left(\pi/2,\sqrt{3}\pi/2\right)+\bq$, we have
$$
\xi_{\bk} \simeq q_x+\sqrt{3}q_y,\,
\xi_{\bk+\mathbf{Q}_1} \simeq  q_x-\sqrt{3}q_y,\,
\xi_{-\bk+\mathbf{Q}_2} \simeq  -2q_x,\,
\xi_{-\bk+\mathbf{Q}_3} \simeq \frac{1}{2}(q^2_x-3q^2_y).
$$
For $\bk=\left(-\pi/2,\sqrt{3}\pi/2\right)+\bq$, we have
$$
\xi_{\bk} \simeq -q_x+\sqrt{3}q_y,\,
\xi_{\bk+\mathbf{Q}_1} \simeq  -q_x-\sqrt{3}q_y,\,
\xi_{-\bk+\mathbf{Q}_2} \simeq  \frac{1}{2}(q^2_x-3q^2_y),\,
\xi_{-\bk+\mathbf{Q}_3} \simeq 2q_x.
$$
Thus we can obtain the following approximation of $g^{(4)}_{\theta,2}$,
\begin{equation}
	\begin{aligned}
		g^{(4)}_{\theta,2} &\simeq
		\sum_{\bq} \frac{-\beta(q_x+\sqrt{3}q_y)+\frac{\beta^3(q_x+\sqrt{3}q_y)^3}{12}}{2\sqrt{3}q_y
			\left(-q_x+\sqrt{3}q_y
			\right)
			\left(q_x+\sqrt{3}q_y
			\right)
		}
		+
		\frac{-\beta(-q_x+\sqrt{3}q_y)+\frac{\beta^3(-q_x+\sqrt{3}q_y)^3}{12}}{2\sqrt{3}q_y
			\left(-q_x+\sqrt{3}q_y
			\right)
			\left(q_x+\sqrt{3}q_y
			\right)
		}\\
		&=\sum_{\bq}\frac{\beta}{q^2_x-3q^2_y}-\frac{\beta^3(q^2_x+q^2_y)}{4\left(q^2_x-3q^2_y\right)}.
	\end{aligned}
\end{equation}
Finally, we obtain the following expression,
\begin{equation}
		g^{(4)}_{\theta} =g^{(4)}_{\theta,1} +g^{(4)}_{\theta,2} 
		\simeq \sum_{\bq}\frac{\beta^3\left(q^2_x-3q^2_y\right)}{12\left(q^2_x-3q^2_y\right)}=\sum_{\bq}\frac{\beta^3}{12}>0.
\end{equation}
Hence, we find that $g^{(4)}_{\theta}$ is also positive.

For positive $g^{(4)}_{\phi}$ and $g^{(4)}_{\theta}$, according to Eq.~\eqref{seq:F4ff}, we should minimize the following functions to obtain the minimum free energy up to $\Delta^4$,
\begin{subequations}
	\begin{align}
	h_\phi&=\cos(2\phi_1)+\cos(2\phi_2)+\cos(2\phi_3),\\
	h_\theta&=\cos(2\theta_2-2\theta_1)+\cos(2\theta_3-2\theta_2)+\cos(2\theta_1-2\theta_3).
\end{align}
\end{subequations}
It is easy to see that $h_\phi$ reaches its minimum $-3$ at $\phi_{\alpha}=\pm \pi/2$. 
For $h_\theta$, let $x=\theta_2-\theta_1$ and $y=\theta_3-\theta_2$, we have
$$h_\theta = \cos(2 x) + \cos(2 y) + \cos(2(x+y)).$$
$h_\theta$ reaches its minimum $-3/2$ at $x=y=\pm2\pi/3\,(\mathrm{mod}\,\pi)$. Thus, our PDW state acquires minimum free energy at $\phi=\pm \pi/2$ and $\theta_2-\theta_1=\theta_3-\theta_2=\pm2\pi/3\,(\mathrm{mod}\,\pi)$, leading to a spatial inversion symmetry and TRS breaking state.

\section{\uppercase\expandafter{\romannumeral6}. Local density of states}\label{ssec:LDOS}
	Since there is no spin-flip effect, the charge density of the system is of the following form,
	\begin{equation}
		\rho (\br_i) = \frac{1}{N}\left(\langle c^\dagger_{\br_i,\uparrow}c_{\br_i,\uparrow}\rangle + \langle c^\dagger_{\br_i,\downarrow}c_{\br_i,\downarrow}\rangle \right)=
		\frac{2}{N} \langle c^\dagger_{\br_i,\uparrow}c_{\br_i,\uparrow}\rangle.
	\end{equation}
	where $c^{\dagger}_{\br_i,\uparrow(\downarrow)}$ is fermion creation operator at site $\br_i$ with spin $\uparrow(\downarrow)$. $\langle \rangle$ denotes the expectation value. By performing the Fourier transformation $c_{\br_i,\sigma}=\frac{1}{\sqrt{N}} \sum_{\bk}e^{i\bk\cdot\br_i}c_{\bk,\sigma}$, we obtain the charge density in $\bq$ space,
	\begin{equation}
		\rho (\bq) = \sum_{\br_i} e^{-i\bq \cdot \br_i}\rho(\br_i)
		= \frac{2}{N} \sum_{\br_i}e^{-i\bq\cdot\br_i} \langle c^\dagger_{\br_i,\uparrow}c_{\br_i,\uparrow}\rangle
		= \frac{2}{N^2}\sum_{\br_i}\sum_{\bk, \bk'}e^{-i(\bk-\bk'+\bq) \cdot\br_i}\langle c^\dagger_{\bk,\uparrow}c_{\bk',\uparrow}\rangle
		= \frac{2}{N}\sum_{\bk}\langle c^\dagger_{\bk,\uparrow}c_{\bk + \bq,\uparrow} \rangle.
	\end{equation}
	Using Eqs.~(5) in the main text, $\rho (\bq)$ can be expressed in terms of Bogoliubov quasi-particle $\gamma_{\bk, \sigma,i}$,
	\begin{align}
		\begin{split}
			\rho (\bq) =& \frac{2}{N}\sum_{\bk}\sum_{j, l}
			\left\langle
			\left(
			u_{1j}(\bk)\gamma^{\dagger}_{\bk, \uparrow, j} + v_{1j}(\bk)\gamma_{-\bk, \downarrow, j}
			\right)
			\left(
			u_{1l}^*(\bk+\bq)\gamma_{\bk+\bq, \uparrow, l} + v_{1l}^*(\bk+\bq)\gamma^{\dagger}_{-\bk-\bq, \downarrow, l}
			\right)
			\right\rangle \\
			=&  \frac{2}{N}\sum_{\bk}\sum_{j, l}
			u_{1j}(\bk)u_{1l}^*(\bk+\bq)
			\left\langle \gamma^{\dagger}_{\bk, \uparrow, j}\gamma_{\bk+\bq, \uparrow, l}
			\right\rangle
			+
			v_{1j}(\bk)v_{1l}^*(\bk+ \bq)
			\left\langle \gamma_{-\bk, \downarrow, j}\gamma^{\dagger}_{-\bk-\bq, \downarrow, l}
			\right\rangle \\
			=&  \frac{2}{N}\sum_{\bk}\sum_{j}
			\left[
			u_{1j}(\bk)u_{1j}^*(\bk+\bq)
			\left\langle \gamma^{\dagger}_{\bk, \uparrow, j}\gamma_{\bk, \uparrow, j}
			\right\rangle
			+
			v_{1j}(\bk)v_{1j}^*(\bk+ \bq)
			\left\langle \gamma_{-\bk, \downarrow, j}\gamma^{\dagger}_{-\bk, \downarrow, j}
			\right\rangle
			\right] \bar{\delta}_{\bk, \bk + \bq} \\
			=&  \frac{2}{N}\sum_{\bk}\sum_{j}
			\left[
			u_{1j}(\bk)u_{1j}^*(\bk+\bq)
			n_F\left(E(\bk)^{+}_j \right)
			+
			v_{1j}(\bk)v_{1j}^*(\bk+ \bq)
			n_F\left(E(\bk)^{-}_j \right)
			\right] \bar{\delta}_{\bk, \bk + \bq},
		\end{split}
	\end{align}
	where we use $\left\langle \gamma^{\dagger}_{\bk, \uparrow, j}\gamma_{\bk', \uparrow, l}
	\right\rangle = n_F\left(E(\bk)^{+}_j\right) \bar{\delta}_{\bk,\bk'}\delta_{j, l}$ and
	$\left\langle \gamma_{-\bk, \downarrow, j}\gamma^{\dagger}_{-\bk', \downarrow, l}
	\right\rangle = n_F\left(E(\bk)^{-}_j\right) \bar{\delta}_{\bk,\bk'}\delta_{j, l}$. Here
	$$
	\bar{\delta}_{\bk, \bk'} =
	\left\{
	\begin{aligned}
		&1,\quad \bk - \bk' = m \mathbf{Q}_1 + n \mathbf{Q}_2, m,n \in \mathcal {Z}\\
		&0,\quad \text{otherwise}\\
	\end{aligned} \right.
	$$
	Thus, the Fourier transformation of the LDOS, $\rho(\bq, \omega)$ reads
	\begin{equation}\label{seq:tri-LDOS0}
		\rho (\bq, \omega)
		=  -\frac{2}{N}\sum_{\bk}\sum_{j}
		\left[
		u(\bk)_{1j}u^*(\bk+\bq)_{1j}\frac{\partial n_{F}\left(\omega-E(\bk)^{+}_j\right)}{\partial \omega}+
		v(\bk)_{1j}v^*(\bk+\bq)_{1j}\frac{\partial n_{F}\left(\omega-E(\bk)^{-}_j\right)}{\partial \omega}
		\right] \bar{\delta}_{\bk, \bk + \bq}.
	\end{equation}
	Using the folding property, Eq.~\eqref{seq:tri-LDOS0} can be written into the following form,
	\begin{equation}\label{seq:tri-LDOS1}
		\rho (\bq, \omega)
		=  -\frac{1}{8N}\sum_{\bk}\sum_{i,j=1}^{16}
		\left[
		u(\bk)_{ij}u^*(\bk+\bq)_{ij}\frac{\partial n_{F}\left(\omega-E(\bk)^{+}_j\right)}{\partial \omega}+
		v(\bk)_{ij}v^*(\bk+\bq)_{ij}\frac{\partial n_{F}\left(\omega-E(\bk)^{-}_j\right)}{\partial \omega}
		\right] \bar{\delta}_{\bk, \bk + \bq}.
	\end{equation}
	We can see from Eq.~\eqref{seq:tri-LDOS1} that $\bq=m\mathbf{Q}_1+n\mathbf{Q}_2$ ($m,n \in \mathcal {Z}$) is necessary for $\rho(\bq,\omega)$ being nonzero. Taking into account the symmetry of the system and $\rho(\bq, \omega) = \rho^*(-\bq, \omega)$ , there are four possible nonzero $|\rho(\bq, \omega)|$ in the BZ,
	\begin{subequations}
		\begin{align}
			\rho_A(\omega)&=|\rho(\bq=\mathbf{0}, \omega)|,\\
			\rho_B(\omega)&=|\rho(\bq =\pm \mathbf{Q}_1, \omega)|=|\rho(\bq= \pm \mathbf{Q}_2, \omega)|=
			|\rho(\bq=\pm \mathbf{Q}_3, \omega)|,\\
			\rho_C(\omega)&=|\rho(\bq =\pm 2\mathbf{Q}_1, \omega)|=|\rho(\bq= \pm 2\mathbf{Q}_2, \omega)|=
			|\rho(\bq=\pm 2\mathbf{Q}_3, \omega)|,\\
			\rho_D(\omega)&=|\rho(\bq=\pm (\mathbf{Q}_1-\mathbf{Q}_2), \omega)|=	|\rho(\bq=\pm (\mathbf{Q}_2-\mathbf{Q}_3), \omega)|=
			|\rho(\bq=\pm (\mathbf{Q}_3- \mathbf{Q}_1), \omega)|.
		\end{align}
	\end{subequations}
	$\rho_A(\omega)$ is just the DOS $\rho(\omega)$ we calculated in the main text.
	
\section{\uppercase\expandafter{\romannumeral7}. Spontaneous loop current at ground state}\label{ssec:loop-current}	
The following formula of nearest neighbor current $J_{i,j}$ is adopted to study the loop current at ground state,
	\begin{align*}
	J_{i, j}=&\sum_{\sigma}\left\langle
	i\left(
	c^{\dagger}_{\br_i,\sigma}c_{\br_j,\sigma}-c^{\dagger}_{\br_j,\sigma}c_{\br_i,\sigma}
	\right)
	 \right\rangle \\
	=& \frac{2i}{N}\sum_{\bk,\bk'}\left(
	e^{-i\bk\cdot\br_i+i\bk'\cdot\br_j}
	-e^{-i\bk\cdot\br_j+i\bk'\cdot\br_i}
	\right)
	\left\langle
	c^{\dagger}_{\bk,\uparrow}c_{\bk',\uparrow}
	\right\rangle\\
	=&\frac{2i}{N}\sum_{\bk,\bq}e^{i\bq\cdot\br_i}\left(
	e^{i(\bk+\bq)\cdot\delta\br}
	-e^{-i\bk\cdot\delta\br}
	\right)
	\left\langle
	c^{\dagger}_{\bk,\uparrow}c_{\bk+\bq,\uparrow}
	\right\rangle,
\end{align*}
where $\delta\br=\br_j-\br_i$.

As we do in the calculation of LDOS above, $J_{i,j}$  can be expressed in terms of Bogoliubov quasi-particle $\gamma_{\bk, \sigma,i}$ according to Eqs.~(5),
\begin{align}\label{seq:loop-current}
	\begin{split}
		J_{i,j} =\frac{2i}{N}\sum_{\bk,\bq}
		e^{i\bq\cdot\br_i}\left(
		e^{i(\bk+\bq)\cdot\delta\br}
		-e^{-i\bk\cdot\delta\br}
		\right)
		\sum_{l}
		\left[
		u_{1l}(\bk)u_{1l}^*(\bk+\bq)
		n_F\left(E(\bk)^{+}_l \right)
		+
		v_{1l}(\bk)v_{1l}^*(\bk+ \bq)
		n_F\left(E(\bk)^{-}_l \right)
		\right] \bar{\delta}_{\bk, \bk + \bq},
	\end{split}
\end{align}
for a given $\bk$, there are 16 $\bq$ in the first BZ that can give a nonzero contribution to the summation $\sum_{\bq}$, according to the factor $\bar{\delta}_{\bk,\bk+\bq}$.

Using Eq.~\eqref{seq:loop-current}, we calculate the loop current in the $4\times 4$ enlarged unit cell at $\frac{k_B T}{\Delta}=\frac{1}{10}$, the result is shown in Fig.~\ref{sfig:loop-current}. We can see from Fig.~\ref{sfig:loop-current} that the ground state shows spontaneous loop current.	
\begin{figure}[hb]
	\includegraphics[width=\textwidth]{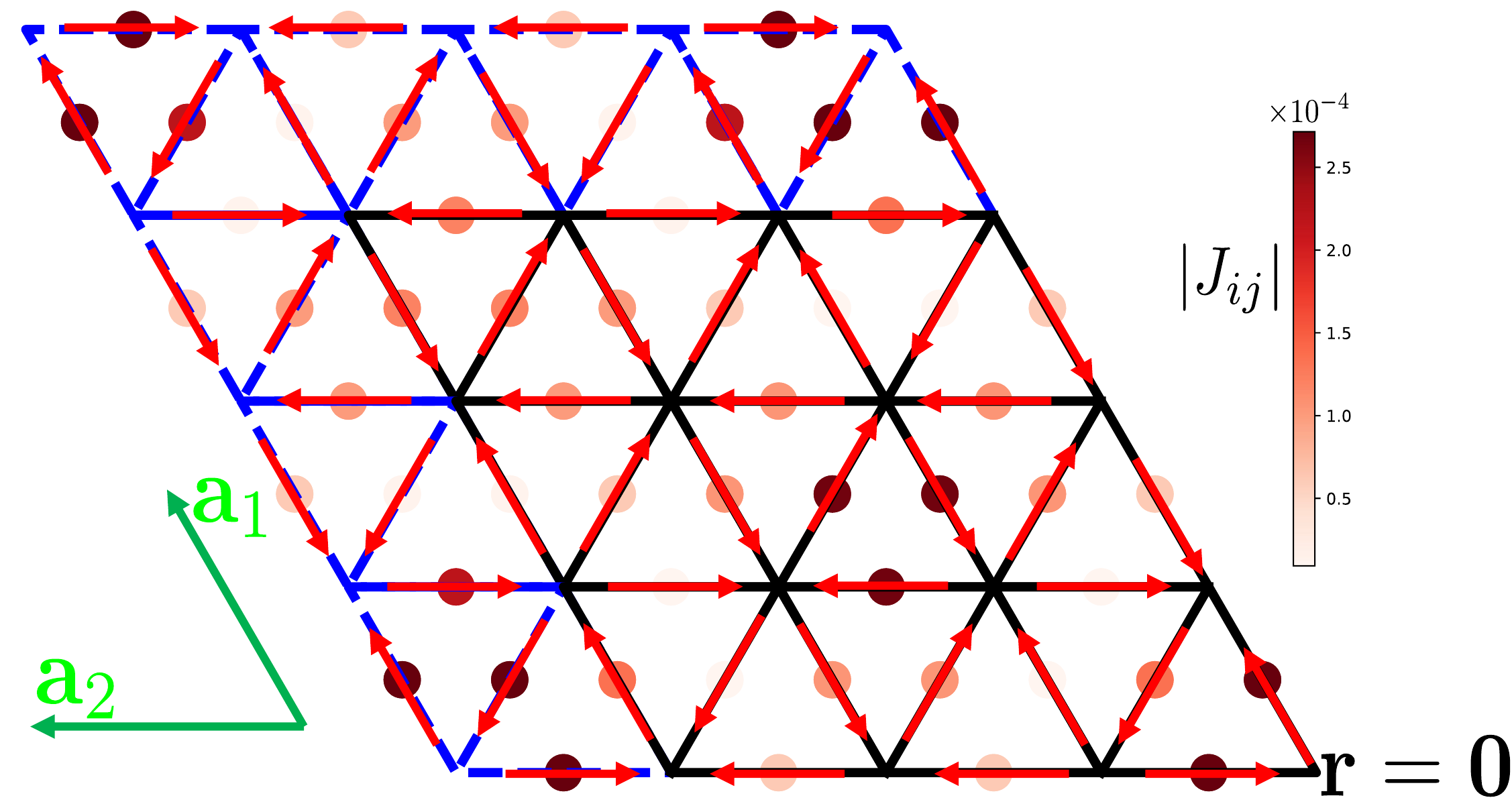}
	\caption{Spontaneous loop current $|J_{i,j}|$ at ground state at $k_BT/\Delta = 1/10$. The red arrows show the directions of currents and the circles at the corresponding links represent the intensity. Here $\Lambda=0.1$, $\Delta=0.02$, $\phi_\alpha=\pi/2$, $\theta_1=0$, $\theta_2=2\pi/3$ and $\theta_3=-2\pi/3$ have been chosen.}\label{sfig:loop-current}
\end{figure}

\end{widetext}

\end{document}